\documentclass[12pt]{article}
\usepackage[colorlinks=true,backref=true,linkcolor=blue,anchorcolor=black,citecolor=blue,filecolor=black,menucolor=black,pagecolor=black,urlcolor=blue]{hyperref}

\usepackage{amsmath}
\usepackage{mathrsfs}
\usepackage[Symbol]{upgreek}
\usepackage{bbm}
\usepackage{dsfont}
\usepackage{amssymb}
\usepackage[vcentermath]{youngtab}
\usepackage{caption}
\usepackage{graphicx}
\usepackage{wasysym}

\usepackage{verbatim}
\usepackage{accents}

\usepackage{tikz}
\usepackage{mathtext}
\usepackage[T1]{fontenc}
\usepackage[english]{babel}
\usepackage{tabularx}
\usepackage{longtable}
\usepackage{subfigure}
\usepackage{setspace}
\usepackage{amsthm}

\newcommand{\rk}{\textrm{rk}}

\newcommand{\be}{\begin{eqnarray}}
\newcommand{\ee}{\end{eqnarray}}
\newcommand{\nn}{\nonumber}

\newcommand{\id}{\text{id}}

\begin{document}
\selectlanguage{english}
\allowdisplaybreaks[1]
\renewcommand{\thefootnote}{\fnsymbol{footnote}}
\numberwithin{equation}{section}
\def\corr{$\spadesuit \, $}
\def\trefle{$ \, $}
\def\kscorr{$\diamondsuit \, $}
\begin{titlepage}
\begin{flushright}
\
\vskip -2.5cm
{\small AEI-2012-035}\\
\vskip 1cm
\end{flushright}
\begin{center}
{\Large \bf
Eisenstein series for infinite-dimensional\\[4mm] U-duality groups}
\lineskip .75em
\vskip20mm
\normalsize
{\large  Philipp Fleig${}^{1,2,3}$, Axel Kleinschmidt${}^{1,4}$}

\vskip 1 em
${}^1${\it Max-Planck-Institut f\"{u}r Gravitationsphysik (Albert-Einstein-Institut)\\
Am M\"{u}hlenberg 1, DE-14476 Potsdam, Germany}
\vskip 1 em
${}^2${\it Freie Universit\"at Berlin, Institut f\"ur Theoretische Physik,\\
Arnimallee 14, 14195 Berlin, Germany}
\vskip 1 em
${}^3${\it Universit\'e de Nice-Sophia Antipolis\\
Parc Valrose, FR-06108 Nice Cedex 2, France}
\vskip 1 em
${}^4${\it International Solvay Institutes\\
ULB-Campus Plaine CP231, BE-1050 Brussels, Belgium}

\vskip20mm

\end{center}

\begin{abstract}
{\footnotesize
We consider Eisenstein series appearing as coefficients of curvature corrections in the low-energy expansion of type II string theory four-graviton scattering amplitudes. We define these Eisenstein series over all groups in the $E_n$ series of string duality groups, and in particular for the infinite-dimensional Kac--Moody groups $E_9$, $E_{10}$ and $E_{11}$. We show that, remarkably, the so-called constant term of Kac--Moody-Eisenstein series contains only a finite number of terms for particular choices of a parameter appearing in the definition of the series. This resonates with the idea that the constant term of the Eisenstein series encodes perturbative string corrections in BPS-protected sectors allowing only a finite number of corrections. We underpin our findings with an extensive discussion of physical degeneration limits  in $D<3$ space-time dimensions.}
\end{abstract}

\end{titlepage}
\renewcommand{\thefootnote}{\arabic{footnote}}
\setcounter{footnote}{0}

\tableofcontents


\section{Introduction}
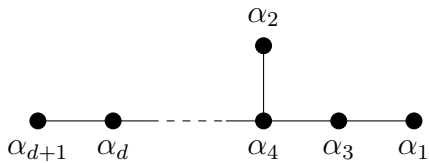
\begin{figure}[t]
\centering
\begin{tikzpicture}
[place/.style={circle,draw=black,fill=black,
inner sep=0pt,minimum size=6}]
\draw (0,0) -- (1,0);
\draw (1,0) -- (1.5,0);
\draw[dashed] (1.5,0) -- (2.5,0);
\draw (2.5,0) -- (3,0);
\draw (3,0) -- (3,1);
\draw (3,0) -- (4,0);
\draw (4,0) -- (5,0);
\node at (0,0) [place,label=below:$\alpha_{d+1}$] {};
\node at (1,0) [place,label=below:$\alpha_{d}$] {};
\node at (3,0) [place,label=below:$\alpha_4$] {};
\node at (4,0) [place,label=below:$\alpha_3$] {};
\node at (5,0) [place,label=below:$\alpha_1$] {};
\node at (3,1) [place,label=above:$\alpha_2$] {};
\end{tikzpicture} 
\caption{\em Dynkin diagram for $E_{d+1}$. \label{fig:Edplus1Diag}}
\end{figure}

Over the last 15 years, a lot of work has been devoted to understanding dualities in string theory. Dualities are discrete symmetries under which string theory is invariant. For toroidal compactifications of type IIB string theory from ten down to $D=10-d$ space-time dimensions on a $d$-dimensional torus $T^d$ these duality symmetries are thought to be contained in the continuous symmetries of the (maximal) low energy supergravities. These are given by the split real groups $E_{d+1}(\mathbb{R})$ for $d\leq 8$ \cite{Cremmer:1978ds,Cremmer:1979up,Nicolai:1987kz,Cremmer:1997ct} and summarised in the first column of Table~\ref{dualitygroups}. Following~\cite{Julia:1980gr,Julia:1982gx}, we have joined conjectural rows for $D=0,1$ to the table.The (conjectured) duality symmetries~\cite{HullTownsend,Font:1990gx} are listed in the last column; they are the corresponding Chevalley groups. The Dynkin diagram corresponding to the various groups is given in Figure~\ref{fig:Edplus1Diag}.
\begin{table}[t!]
\begin{center}
\begin{tabular}{ | c || c  c  c  | }
  \hline                       
  $D$ & $E_{d+1}(\mathbb{R})$ & $K(E_{d+1})$ & $E_{d+1}(\mathbb{Z})$ \\ \hline \hline
 $10$ & $SL(2,\mathbb{R})$ & $SO(2)$ & $SL(2,\mathbb{Z})$ \\ \hline
 $9$ & $\mathbb{R}^+\times SL(2,\mathbb{R})$ & $SO(2)$ & $ SL(2,\mathbb{Z})$ \\ \hline
 $8$ & $SL(2,\mathbb{R})\times SL(3,\mathbb{R})$ & $SO(3)\times SO(2)$ & $SL(2,\mathbb{Z})\times SL(3,\mathbb{Z})$ \\ \hline
 $7$ & $SL(5,\mathbb{R})$ & $SO(5)$ & $SL(5,\mathbb{Z})$  \\ \hline
 $6$ & $SO(5,5,\mathbb{R})$ & $SO(5) \times SO(5)$ & $SO(5,5,\mathbb{Z})$ \\ \hline
 $5$ & $E_{6}(\mathbb{R})$ & $USp(8)$ & $E_{6}(\mathbb{Z})$  \\ \hline
 $4$ & $E_{7}(\mathbb{R})$ & $SU(8)/\mathbb{Z}_2$ & $E_{7}(\mathbb{Z})$ \\ \hline
 $3$ & $E_{8}(\mathbb{R})$ & $Spin(16)/\mathbb{Z}_2$ & $E_{8}(\mathbb{Z})$ \\ \hline
 $2$ & $E_{9}(\mathbb{R})$ & $K(E_{9}(\mathbb{R}))$ & $E_{9}(\mathbb{Z})$ \\ \hline
 $1$ & $E_{10}(\mathbb{R})$ & $K(E_{10}(\mathbb{R}))$ & $E_{10}(\mathbb{Z})$ \\ \hline
  $0$ & $E_{11}(\mathbb{R})$ & $K(E_{11}(\mathbb{R}))$ & $E_{11}(\mathbb{Z})$ \\ \hline
\end{tabular}
\caption{\label{dualitygroups}\em List of the split real forms of the duality groups one obtains when compactifying type IIB string theory on a $d$-torus to $D=10-d$ space-time dimensions. We also list the corresponding maximal compact subgroups and the last column contains the discrete versions, which appear in string theory. The last two rows are conjectural as are the corresponding discrete groups for $D\leq 3$. Note that $E_{10}$ and $E_{11}$ as appearing here are thought of as symmetries of the toroidally compactified theory; in contrast to the farther-reaching conjectures of~\cite{Damour:2002cu} and~\cite{West:2001as}.}
\end{center}
\end{table}

One way in which the invariance of type IIB string theory under the groups shown in Table \ref{dualitygroups} becomes manifest is in string scattering amplitudes. The scattering amplitudes depend on the moduli $\Phi\in \mathcal{M}_{d+1}$ of the compactified theory given by the coset $E_{d+1}/K_{d+1}$. Here $K_{d+1}=K(E_{d+1})$ is the maximal compact subgroup of $E_{d+1}$ and a complete list is also given in Table \ref{dualitygroups}. The scattering amplitude is then invariant under the discrete group $E_{d+1}({\mathbb{Z}})$; it transforms as an automorphic function under it.\footnote{It has also been suggested that so-called transforming automorphic forms  that are not invariant but covariant under duality play a r\^ole for correction terms~\cite{Green:1997me,Green:2003an,Lambert:2006ny,Bao:2007er,Michel:2007vh}.}  Starting with the work of \cite{GutperleGreen} it has been shown in~\cite{Green:1997di,Green:1997as,Green:1997me,Kiritsis:1997em,Antoniadis:1997zt,Pioline:1998mn,Green:1998by,Obers:1999um,Basu:2006cs,Basu:2007ru,Green:2006gt} that when considering four-graviton scattering, one can make precise statements about the form of the three lowest orders in the low-energy expansion of the four-graviton scattering amplitude. We will now, following loosely \cite{GreenAutoProps,Pioline:2010kb,GreenESeries},  briefly present some of the background to our work. 

The expansion at low energies of the four-graviton scattering amplitude $\mathcal{A}^D(s,t,u)$ in $D$ space-time dimensions is a function of the Mandelstam variables $s$, $t$ and $u$. It can be written as a sum $\mathcal{A}_{\text{analytic}}^D+\mathcal{A}_{\text{non-analytic}}^D$, with the first term being an analytic function of the Mandelstam variables and the second term being non-analytic in these variables~\cite{Green:1997as}. In the present work we will mainly focus on the analytic part $\mathcal{A}_{\text{analytic}}^D(s,t,u)$. The non-analytic contribution also plays a r\^ole in the analysis and we will provide some more comments on this term later on. The analytic part in the expansion takes the form\footnote{The first term on the right-hand-side of this expansion is the classical supergravity tree-level term, determined by the Einstein-Hilbert action. The function $\mathcal{E}^D_{(0,-1)}=3$.}
\begin{align}
\label{FourGravAn}
\mathcal{A}^D_{\text{analytic}}(s,t,u)=\mathcal{E}_{(0,-1)}^{D}(\Phi)\frac{\mathcal{R}^4}{\sigma_3}+\sum_{p=0}^\infty\sum_{q=0}^\infty\mathcal{E}_{(p,q)}^{D}(\Phi)\sigma_2^p\sigma_3^q\mathcal{R}^4\,.
\end{align}
Here, $\sigma_n=(s^n+t^n+u^n)\frac{\ell_D^{2n}}{4^n}$ is a dimensionless combination of the Mandelstam variables, with $\ell_D$ being the Planck length in $D$ space-time dimensions, and $\mathcal{R}^4$ denotes a contraction of four Riemann tensors with a standard 16-index tensor.\footnote{The 16-index tensor is the $t_8t_8$ tensor, which can be found for example in~\cite{Gross:1986iv}.} The $\mathcal{E}_{(p,q)}^{D}(\Phi)$ are automorphic functions of the moduli $\Phi\in\mathcal{M}_{d+1}$. The superscript $D$ indicates that $\mathcal{E}$ is an automorphic function under the duality group in $D=10-d$ space-time dimensions, i.e. $E_{d+1}$. The orders $2p+3q\leq3$, with positive integers $p$ and $q$, have been studied extensively in the literature and a considerable amount of evidence for their precise form has been accumulated in $D\geq 3$. When translated into the effective action, a term of the form $\mathcal{E}_{(p,q)}^{D}(\Phi)\sigma_2^p\sigma_3^q\mathcal{R}^4$ in the scattering amplitude corresponds to a term which is of the form $\mathcal{E}_{(p,q)}^{D}(\Phi)\partial^{2(2p+3q)}\mathcal{R}^4$.

In the present work, we will mainly be concerned with the two lowest orders of string theory corrections in the effective action, namely $\mathcal{E}^D_{(0,0)}\mathcal{R}^4$ and $\mathcal{E}^D_{(1,0)}\partial^4\mathcal{R}^4$, where for simplicity of notation we have dropped the explicit moduli dependence. It has been found that for the low-energy expansion of four-graviton scattering in type IIB superstring theory in $D\geq 3$, $\mathcal{E}^D_{(0,0)}$ and $\mathcal{E}^D_{(1,0)}$ are given by Eisenstein series, multiplied by a suitable normalisation factor\footnote{We do not consider the cases $6\leq D\leq 9$ where the functions are given by sums of Eisenstein series.}~\cite{GreenAutoProps,GreenESeries}
\begin{align}\label{ESeries}
\mathcal{E}^D_{(0,0)}=2\zeta(3)E_{1;3/2}^G,\quad\text{and}\quad\mathcal{E}^D_{(1,0)}=\zeta(5)E_{1;5/2}^G\,,
\end{align}
where $\zeta$ is the Riemann-Zeta function. These Eisenstein series are of the general form $E^G_{i_*;s}$, where $G=E_{d+1}(\mathbb{Z})$ is the duality in group in $D\geq 3$ space-time dimensions under which the Eisenstein series are invariant. The label $i_*$ indicates  a particular chosen simple root of $E_{d+1}$ (see Fig.~\ref{fig:Edplus1Diag})  and $s$ is a generically complex parameter, which enters in the definition of the series. The notation used here will be explained in more detail later. It is interesting to also consider the Fourier expansion of such Eisenstein series, since it allows one to give a physical interpretation of the different terms. In such an expansion one will find two types of terms, which differ in their mathematical structure and physical interpretations. The first type is generally referred to as \textit{the constant term}~\cite{Langlands}. The physical interpretation ascribed to this type is that each of its terms encodes a perturbative (string or M-theory) correction of a certain loop order in the scattering process. For finite-dimensional groups, the number of constant terms is bounded by the order of the corresponding Weyl group and therefore finite. This corresponds to a finite number of perturbative corrections (irrespective of supersymmetry). For the actual series (\ref{ESeries}) occurring in string theory, there are many additional cancellations and the number of constant terms is further reduced drastically; this can be seen as a consequence of the large supersymmetry and its associated non-renormalisation properties~\cite{GutperleGreen}.
The second type of term which appears in the expansion of the Eisenstein series is generally associated with non-perturbative effects, or more precisely instanton corrections, see for example~\cite{GutperleGreen,Green:1997as,Kiritsis:1997em,Obers:1999um,Green:2006gt}. This type of term is often called (abelian or non-abelian) Fourier coefficients.

As proven in~\cite{Green:1998by} for $D\geq 3$, the coefficient functions $\mathcal{E}^D_{(0,0)}$, $\mathcal{E}^D_{(1,0)}$ and $\mathcal{E}^D_{(0,1)}$ of the lowest three orders of curvature corrections each satisfy a Laplace eigenvalue equation defined by the $E_{d+1}$ invariant Laplace operator $\Delta^D$ on the moduli space $\mathcal{M}_{d+1}$ in $D=10-d$ dimensions. In the first two cases this Laplace eigenvalue equation is homogeneous (with source terms only in dimensions when there is a known divergence). For the third case $\mathcal{E}^D_{(0,1)}$, the coefficient of the $\partial^6\mathcal{R}^4$ correction, the equation is always inhomogeneous, where the inhomogeneous term is given by $(\mathcal{E}^D_{(0,0)})^2$. We will give explicit expressions for these Laplace equations later on, supplemented by some further discussion.  Duality invariance and the eigenvalue of the Laplace operator in $D=4$ for various terms has also played a r\^ole in recent discussions of the finiteness (or not) of $N=8$ supergravity, see e.g.~\cite{Bossard:2010dq,Bossard:2010bd,Beisert:2010jx,Kallosh:2011dp,Bossard:2011ij} and references therein.

The functions (\ref{ESeries}) are subject to a number of strong consistency requirements~\cite{GreenAutoProps,Pioline:2010kb} that arise from the interplay of string theory in various dimensions. The consistency conditions are typically phrased in terms of three limits, corresponding to different combinations of the torus radii\footnote{in appropriate units of Planck or string length in the relevant dimensions} and the string coupling becoming large. The three standard limits correspond to (i) decompactification from $D$ to $D+1$ dimensions, (ii) string perturbation theory and (iii) the M-theory limit. In terms of the $E_{d+1}$ diagram this means singling out the nodes $d+1$, $1$ or $2$, respectively, in the three cases. In or above $D=3$ (i.e., up to and including $E_8$), the functions in (\ref{ESeries}) have been successfully subjected to the consistency requirements~\cite{GreenESeries,Pioline:2010kb,GreenSmallRep}. There are also direct checks of their correctness for some dimensions and parts of their expansions (see~\cite{GutperleGreen,Green:1997as,Green:2006gt} and references therein) and general considerations on perturbative expansions for functions constructed from lattice sums (not necessarily satisfying a Laplace equation)~\cite{Lambert:2010pj}. We will provide a heuristic derivation of the parameters entering (\ref{ESeries}) below.

It is natural to ask the question whether these results also extend to the case of the infinite-dimensional symmetry groups $E_9$, $E_{10}$ and $E_{11}$ and their associate duality symmetries $E_9(\mathbb{Z})$, $E_{10}(\mathbb{Z})$ and $E_{11}(\mathbb{Z})$.\footnote{By abuse of notation we will refer to the discrete duality groups $E_n(\mathbb{Z})$ as finite-dimensional duality groups for $n\leq 8$ and as infinite-dimensional duality groups for $n>8$. This sloppiness of terminology helps to make many statements more readable and we will similarly sometimes omit the `$\mathbb{Z}$' in $E_n(\mathbb{Z})$ when it is implicit from the context.} The present work is an attempt to answer this question and provide insights into the technical details of how one can extend the analysis to the infinite-dimensional cases. Issues regarding the physical definition of the charges of states and space-time dependence of the moduli fields in low space-time dimensions will not be addressed. This can be (partly) justified by regarding the Eisenstein series for $E_n$ ($n\geq 9$) as unifying objects that give rise to the more physical series for $n\leq 8$ in special limits like the ones to be discussed in section~\ref{sec:computations}.

A central role in our analysis is played by the precise structure of the constant term of the Eisenstein series shown in \eqref{ESeries}, when $D=2$ and $1$, i.e. for the invariance groups $E_9$ and $E_{10}$. We will also study the constant terms of the $E_{11}$ Eisenstein series.\footnote{The remaining parts of the Fourier expansion are not addressed in our work. Although we assume that there will be a connection to minimal and next-to-minimal representations as in~\cite{Kazhdan:2001nx,Pioline:2010kb,GreenSmallRep} we do not explore this here. A new feature that should arise in for $D<3$ are instanton corrections of objects that are more non-perturbative than NS$5$ branes, i.e., have a string frame `tension' scaling with $g_{\text s}^{\alpha}$ with $\alpha<-2$~\cite{Elitzur:1997zn,Blau:1997du}. Half-BPS states are expected to fill out infinite duality multiplets~\cite{Elitzur:1997zn,ObersUdualMTheory,Englert:2007qb}.} In particular, as will be explained in more detail later, it is not \textit{a priori} clear that the constant terms of these series will be made up of a finite number of terms since now the Weyl groups are of infinite order. However, due to the physical interpretation of the constant term as encoding a \textit{finite} number of perturbative corrections, it is crucial for consistency that the constant terms of the Eisenstein series invariant under infinite-dimensional groups, also only consist of a finite number of terms. Using a technical argument we will show that for special choices of the parameter $s$ appearing in the definition of the Eisenstein series, this is indeed the case. This requirement leaves only a small subset of values of $s$ out of an initially infinite range. These include the ones that appear in the coefficients of the curvature corrections. In this sense restrictive nature of supersymmetry and the infinite-dimensional duality groups is revealed.

The plan of the article is as follows. In section~\ref{defsec} of the paper we will recapitulate the definition of an Eisenstein series over finite-dimensional groups and introduce some of the concepts required to do so. We will then go on to extend this definition to the case of the infinite-dimensional groups and in particular to the affine groups based on work by Garland. Section~\ref{sec:ConstFin} will be concerned with the structure of the constant term which appears in a Fourier-like expansion of the Eisenstein series. In section~\ref{sec:CT} we will give an expression for the constant term of Eisenstein series over affine groups and the results of section~\ref{sec:ConstFin} will be extended to the infinite-dimensional case. We will show that it is possible that the constant term of an Eisenstein series invariant under an affine group contains only finitely many terms for special values of $s$. Section~\ref{sec:computations} contains many of the explicit computations which were carried out, others have been relegated to appendices. In particular, we provide constant terms of the $E_9$-and $E_{10}$-Eisenstein series in three different (maximal) parabolic subgroups. These correspond to three physical degeneration limits that we discuss carefully since the $D=2$ case affords physical and technical novelties.  The results of this paper were announced in~\cite{Talk}.

\section{Definition of Eisenstein series}
\label{defsec}

Before we begin, let us fix some general notation used in this paper. We denote the Lie algebra of a group $G$ by $\mathfrak{g}$ and the set of roots of the algebra is denoted by $\Delta$.  A basis of $\Delta$ is given by the choice of a set of simple roots $\Pi$ and the number of elements in $\Pi$ is equal to the rank of $\mathfrak{g}$. An element of $\Pi$ is generally denoted by $\alpha_i$, where $i=1,\ldots,\rk(\mathfrak{g})$. We also denote the set of positive and negative roots by $\Delta_+$ and $\Delta_-$, respectively. The Cartan subalgebra of $\mathfrak{g}$ is denoted by $\mathfrak{a}$.

In the present article we shall mainly be concerned with the groups of the $E_n$ series of the Cartan classification with $n=1,...,8$, and their infinite-dimensional Kac--Moody extensions for $n>8$, as given by the Dynkin diagram in Fig.~\ref{fig:Edplus1Diag}. More precisely, we are interested in the split real form of these groups, commonly denoted by $E_{n(n)}$. For simplicity of notation, we however denote the split real form simply by $E_n$. As our discussion is aimed at the $E_n$ series, we will state our results for simple and simply-laced algebras.

As already explained in the introduction, the duality groups appearing in reductions of type II string theory are discrete versions of the $E_{d+1}$ groups, which we will denote by $E_{d+1}(\mathbb{Z})$ and take to be the associated Chevalley groups~\cite{Mizoguchi:1999fu,GarlLoop}. These can be thought of as being generated by the integer exponentials of the (real root) generators of $E_{d+1}$  in the Chevalley basis.

\subsection{Borel and parabolic subalgebras}

The {\em Borel subalgebra} $\mathfrak{b}$ of an algebra $\mathfrak{g}$ is defined as
\begin{align}
\mathfrak{b}=\mathfrak{a}\oplus\bigoplus_{\alpha\in\Delta_+}\mathfrak{g}_\alpha\,.
\end{align}
A (standard) {\em parabolic subalgebra} $\mathfrak{p}$  is a  subalgebra of $\mathfrak{g}$ that contains $\mathfrak{b}$. Parabolic subalgebras $\mathfrak{p}$ decompose in general as the direct sum of the so-called Levi subalgebra $\mathfrak{m}$ and the unipotent radical $\mathfrak{n}$
\begin{align}\label{decompparab}
\mathfrak{p}=\mathfrak{m}\oplus\mathfrak{n}\,.
\end{align}
A convenient construction of parabolic subalgebras is obtained by selecting a subset $\Pi_1$ of the set of simple roots $\Pi$. This induces a corresponding subset $\Gamma_1$ of the set of positive roots $\Delta_+$, where the $\Gamma_1$ are those positive roots that are linear combinations of the simple roots in $\Pi_1$ only. The Levi subalgebra and unipotent radical are then defined as
\begin{align}
\mathfrak{m}(\Pi_1)=\mathfrak{a}\oplus\bigoplus_{\alpha\in\Gamma_1\cup-\Gamma_1}\mathfrak{g}_\alpha
\end{align}
and
\begin{align}
\mathfrak{n}(\Pi_1)=\bigoplus_{\substack{\alpha\in\Delta_+\backslash\Gamma_1}}\mathfrak{g}_\alpha
\end{align}
respectively and the parabolic subalgebra is given by
\begin{align}
\mathfrak{p}(\Pi_1)=\mathfrak{a}\oplus\bigoplus_{\alpha\in\Delta_+\cup(-\Gamma_1)}\mathfrak{g}_\alpha = \mathfrak{b}\oplus\bigoplus_{\alpha\in (-\Gamma_1)} \mathfrak{g}_\alpha\,.
\end{align}
There are two types of parabolic subalgebras which are of importance for us. The first is the \textit{minimal parabolic} case, which is obtained, when $\Pi_1=\emptyset$ and corresponds to the Borel subalgebra $\mathfrak{b}$. The second is the \textit{maximal parabolic} case for which $\Pi_1=\Pi\backslash\{\alpha_{i_*}\}$, where $\alpha_{i_*}$ is a (single) simple root.\footnote{Our terminology differs from that used for example in~\cite{GreenESeries} in that there $\Pi_1=\left\{\alpha_{i_*}\right\}$.} Using an abbreviated notation we denote maximal parabolic subalgebras by $\mathfrak{p}_{i_*}$. We will denote the group associated with the subalgebra $\mathfrak{p}_{i_*}$ by $P_{i_*}$. Similar to the decomposition of $\mathfrak{p}$ shown in \eqref{decompparab} we also have
\begin{align}
\label{MaxParDef}
P_{i_*}=M_{i_*}N_{i_*}\,,
\end{align}
where $M_{i_*}$ and $N_{i_*}$ are the groups associated with the subalgebras $\mathfrak{m}_{i_*}$ and $\mathfrak{n}_{i_*}$.\\

\subsection{Eisenstein series over finite-dimensional groups}

Before discussing the definition of Eisenstein series over infinite-dimensional groups, we want to give the definition for the case of a finite-dimensional group $G$. We define the following (Langlands-)Eisenstein series~\cite{Langlands}
\begin{align}
\label{minimalparabfinite}
E^G(\lambda,g)\equiv\sum_{\gamma \in B(\mathbb{Z})\backslash G(\mathbb{Z})}e^{\langle\lambda+\rho|H(\gamma g)\rangle}\,.
\end{align}
where  $G(\mathbb{Z})$ is the Chevalley group of $G$ and $B(\mathbb{Z})=B\cap G(\mathbb{Z})$ the corresponding discrete version of the Borel subgroup $B$. $\lambda$ is a general weight vector of $G$ (not necessarily on the weight lattice) and $\rho$ is the Weyl vector, which is defined as half the sum over all positive roots or alternatively as the sum over all fundamental weights which we will denote by $\Lambda_i$ (with $i=1,\ldots,\rk(G)$). The function $H$ is a map from a general group element $g\in G$ to the Cartan subalgebra $\mathfrak{a}$. Using the standard unique Iwasawa decomposition
\begin{align}
\label{Iwasawa}
G=NAK\,
\end{align}
we write down the action of the map $H$ for a specific group element $g$, decomposed according to \eqref{Iwasawa}
as $g=nak$, as
\begin{align}
\label{Hfunc}
a=e^{H(g)}\,,
\end{align}
This then defines the map $H\,:\,G \to \mathfrak{a}$. The angled brackets in the definition are the standard pairing between the space of weights $\mathfrak{a}^*$ and the Cartan subalgebra $\mathfrak{a}$. We refer to the function defined in (\ref{minimalparabfinite}) as a {\em minimal parabolic Eisenstein series} since it is associated with the minimal parabolic subgroup $B$. The sum (\ref{minimalparabfinite})  converges when the real parts of the inner products $\langle\lambda|\alpha_i\rangle$ for all simple roots $\alpha_i$ are sufficiently large and can be analytically continued to the complexified space of weights except for certain hyperplanes~\cite{Langlands}.

The Eisenstein series (\ref{minimalparabfinite}) is made out of a simple `plane-wave type' function $e^{\langle \lambda+\rho| H(a)\rangle}= e^{w_i \beta^i}$ where $\beta^i$ are coordinates of the Cartan subalgebra and the $w_i$ are constants determined by $\lambda$.\footnote{We use the incorrect terminology `plane wave' with an application to quantum gravity in mind, where the Eisenstein series should describe wavefunctions~\cite{Kleinschmidt:2009cv,Kleinschmidt:2010zz}.} This function is stabilized by the Borel group $B(\mathbb{Z})$ and turned into an automorphic function by summing over all its (inequivalent) $G(\mathbb{Z})$ images determined by $B(\mathbb{Z})\backslash G(\mathbb{Z})$. The plane-wave function is trivially an eigenfunction of the quadratic Laplace operator and all higher-order invariant differential operators. Since all these operators are invariant under the group $G(\mathbb{Z})$ (even $G(\mathbb{R})$), the Eisenstein series $E^G(\lambda,g)$ of (\ref{minimalparabfinite}) is an eigenfunction of all these operators. In particular, its eigenvalue under the $G$-invariant Laplacian $\Delta^{G/K}$ (changing the normalisation of~\cite{GreenESeries}) is
\begin{align}
\Delta^{G/K} E^G(\lambda,g) = \frac12\left( \langle \lambda|\lambda\rangle -\langle \rho|\rho\rangle\right) E^G(\lambda,g)\,.
\end{align}
This, of course, is the same eigenvalue as that of the quadratic Casimir on a representation with highest weight $\Lambda=-(\lambda+\rho)$ up to normalisation. The inner product $\langle\cdot|\cdot\rangle$ is normalised such that $\langle \alpha_i|\alpha_i\rangle=2$ for simple roots $\alpha_i$.

One can consider a special case of the Eisenstein series defined in \eqref{minimalparabfinite} by imposing the additional condition 
\begin{align}
\lambda=2s\Lambda_{i_*}-\rho
\end{align}
for a chosen $i_*$. This condition implies that $\lambda+\rho$ will be orthogonal to all simple roots $\alpha_i$ with $i\neq i_*$. The parameter $s$ which appears here can in general be any complex number. However, we will see later that in the cases which are relevant for us in the context of superstring graviton scattering, $s$ will be purely real and take half-integer values. With a short calculation, see e.g.~\cite{GreenESeries}, one can show that the sum in (\ref{minimalparabfinite}) now becomes a sum over the coset $P_{i_*}(\mathbb{Z})\backslash G(\mathbb{Z})$ and that it takes the form
\begin{align}\label{maxparabfinite}
E^G_{i_*;s}(g):= E^G(2s\Lambda_{i_*}-\rho,g) = \sum_{\gamma \in P_{i_*}(\mathbb{Z})\backslash G(\mathbb{Z})}e^{2s\langle\Lambda_{i_*}|H(\gamma g)\rangle}\,.
\end{align}
Here, $P_{i_*}$ is the maximal parabolic subgroup defined by the node $i_*$. For obvious reasons, this function is referred to as a {\em maximal parabolic Eisenstein series} and $s$ is called the {\em order} of the Eisenstein series. In some cases, equivalent expressions in terms of (restricted) lattice sums exist~\cite{GutperleGreen,Obers:1999um,Pioline:2009qt,Bao:2009fg,Angelantonj:2011br}.

\subsection{Eisenstein series over Kac--Moody groups}
\label{sec:EKM}

The theory of Eisenstein series defined over affine (loop) groups was first developed by Garland and is comprehensively described in \cite{GarlLoop} (see also~\cite{GarlConvergence}). Indeed, the definition of Eisenstein series over affine groups proceeds in much the same way as the one for the finite groups. There are, however, some subtleties which we shall explain in the following. 

A hat is used to denote objects of affine type. Starting from a finite-dimensional, simple and $\mathbb{R}$-split Lie algebra $\mathfrak{g}$ one constructs the non-twisted affine extensions as
\begin{align}
\hat{\mathfrak{g}} = \mathfrak{g}[[t,t^{-1}]]\oplus c \mathbb{R}\oplus d\mathbb{R}\,.
\end{align}
The first summand is the algebra of formal Laurent series over $\mathfrak{g}$ (the loop algebra) and the other summands are the central extension and derivation, respectively. The algebra $\hat{\mathfrak{g}}$ has a Cartan subalgebra of dimension $\dim (\mathfrak{a})+2$ and its roots decompose into real roots and imaginary roots (see \cite{Kac90}). 

The real affine group $\hat{G}$ (in a given representation over $\mathbb{R}$) is defined by taking the closure of exponentials of the {\em real} root generators of the non-twisted affine algebra. Due to the structure of the commutation relations where $d$ never appears on the right hand side, the group thus generated will not use the derivation generator. $\hat{G}$ has the following Iwasawa decomposition
\begin{align}
\hat{G}=\hat{N}\hat{A}\hat{K}
\end{align}
in an analogous way to (\ref{Iwasawa}), but now $\hat{A}$ is the exponential of the $ \dim(\mathfrak{a})+1$ dimensional abelian algebra $\hat{\mathfrak{a}}\equiv\mathfrak{a}\oplus c \mathbb{R}$ only~\cite{GarlLoop}. $\hat{G}$ does not include the group generated by the derivation $d$; we denote by $E_9$ the group $\hat{E_8}$ with the derivation added to it.

Similar to the definition of the Eisenstein series over finite-dimensional groups, in the infinite-dimensional case one can define in a meaningful manner
\begin{align}
\label{ESeriesAffine}
E^{\hat{G}}({\hat{\lambda}};\hat{g},v)=\sum_{\hat{\gamma}\in\hat{B}(\mathbb{Z})\backslash\hat{G}(\mathbb{Z})}e^{\langle\hat{\lambda}+\hat{\rho}|\hat{H}(\hat{\gamma}\,v^d\,\hat{g})\rangle}\,,
\end{align}
where $v$ parameterises the group associated with the derivation generator $d$ and is written as $e^{-r d}$ in~\cite{GarlLoop}. This definition of the Eisenstein series is derived in \cite{GarlLoop} and the convergence of the series is proven for $\textrm{Re}\langle \hat\lambda | \hat{\alpha}_i\rangle>1$ for $i=1,\ldots,\rk(G)+1$. The domain of definition can be extended by meromorphic continuation. One important special property of the affine case that enters in (\ref{ESeriesAffine}) is the definition of the affine Weyl vector $\hat{\rho}$: The usual requirement that the Weyl vector have inner product $\langle\hat{\rho}|\hat{\alpha}_i\rangle=1$ with all affine simple roots $\hat{\alpha}_i$ does not fix $\hat{\rho}$ completely; it is only defined up to shifts by the so-called (primitive) null root $\hat{\delta}$ that has vanishing inner product with all $\hat{\alpha}_i$~\cite{Kac90}. We choose the standard convention that $\hat{\rho}$ is the sum of all the fundamental weights~\cite{Kac90}, i.e., it acts on the derivation $d$ by $\hat{\rho}(d)=0$. Associated with the existence of the null element $\hat{\delta}$ is also the existence of a particular type of a fairly simple automorphic function given $\hat{\lambda} = k \hat{\delta}-\hat{\rho}$ for any $k\in \mathbb{R}$, where $\hat{\delta}$ is the primitive affine null root of the affine root system. This is the automorphic version of the fact that there are infinitely many trivial representations whose characters differ by factors $e^{k\hat{\delta}}$~\cite{Kac90}. We denote these special Eisenstein series by
\begin{align}
\label{affspecial}
\mathcal{A}_k = \sum_{\hat{\gamma}\in\hat{B}(\mathbb{Z})\backslash\hat{G}(\mathbb{Z})}e^{\langle k \hat{\delta}|\hat{H}(\hat{\gamma}\, v^d\,\hat{g})\rangle}
= v^k\,.
\end{align}
More generally, we can always multiply any affine Eisenstein series by an arbitrary power of $v$ and still obtain an Eisenstein series.

As in the finite-dimensional case, the Eisenstein series (\ref{ESeriesAffine}) is an eigenfunction of the full affine Laplacian and has eigenvalue
\begin{align}
\Delta^{\hat{G}/\hat{K}} E^{\hat{G}}(\hat{\lambda};\hat{g},v) = \frac12( \langle \hat{\lambda}|\hat{\lambda}\rangle -\langle \hat{\rho}|\hat{\rho}\rangle) E^{\hat{G}}(\hat{\lambda};\hat{g},v)\,.
\end{align}
The Laplacian itself is not unambiguously defined because of the ambiguity in $\hat{\rho}$ (related to a rescaling of the overall volume of moduli space). We reiterate that we adopt consistently the convention that $\hat{\rho}$ has no $\hat{\delta}$ part. An important difference to the finite-dimensional case is that there are no higher order polynomial invariant differential operators that help to determine $\hat{\lambda}$ but only transcendental ones~\cite{KacPeterson}. We have not investigated their action on (\ref{ESeriesAffine}).

By imposing the additional condition $\hat{\lambda}=2s\hat{\Lambda}_{i_*}-\hat{\rho}$ on the (minimal) Eisenstein series defined above in (\ref{ESeriesAffine}) one can again obtain a maximal parabolic Eisenstein series:
\begin{align}
E^{\hat{G}}_{i_*;s}(\hat{g},v) :=\sum_{\hat{\gamma}\in\hat{P}_{i_*}(\mathbb{Z})\backslash\hat{G}(\mathbb{Z})}e^{2s\langle\hat{\Lambda}_{i_*}|\hat{H}(\hat{\gamma}\, v^d\,\hat{g})\rangle}\,.
\end{align}

Turning to more general Kac--Moody groups, we will assume that the Eisenstein series for $E_n(\mathbb{Z})$ with $n>9$ are defined formally exactly as in (\ref{minimalparabfinite}). A proof for the validity of this formula (i.e. existence via convergence) is not known to our knowledge but for sufficiently large real parts of $\lambda$ one should obtain a convergent bounding integral and then continue meromorphically. The definition of the real group and the Chevalley group proceeds along the same lines as in the affine case~\cite{KacPeterson}. The expression for the Laplace eigenvalue is as before and is unambiguous for $E_n$ with $n>9$. We do not address the issue of square integrability of Eisenstein series for Kac--Moody groups.

\section{Eisenstein series and constant terms}
\label{sec:ConstFin}

We now turn to the analysis of the constant terms of Eisenstein series of the type (\ref{minimalparabfinite}) or (\ref{ESeriesAffine}). In this section we restrict mainly to the finite-dimensional duality groups $G$ and treat the infinite-dimensional case in the next section.

\subsection{Constant term formul\ae}

The constant terms are those terms that do not depend on those $G/K$ coset space coordinates associated with the unipotent part $N$ in (\ref{Iwasawa}) but only on the Cartan subalgebra coordinates. They are hence obtained by integrating out the unipotent part (using the invariant Haar measure):\footnote{We note the similarity with Weyl character formula.}
\begin{align}
\label{minparexp}
\int\limits_{N(\mathbb{Z})\backslash N(\mathbb{R})} E^G(\lambda,g) \mathrm{d}n = \sum_{w\in\mathcal{W}} M(w,\lambda) e^{\langle w\lambda + \rho | H(g)\rangle}\,,
\end{align}
where we have already applied Langlands' constant term formula \cite{Langlands} for evaluating the integral. The constant terms are hence given by a sum over the Weyl group $\mathcal{W}$ of $E_{d+1}$ with individual summands being the numerical factor $M(w,\lambda)$ times a monomial of the Cartan subalgebra coordinates. 
The numerical factors $M(w,\lambda)$ are given explicitly by
\begin{align}
\label{LangCoeffs}
M(w,\lambda)=\prod_{\substack{\alpha \in \Delta_+\\w\alpha \in \Delta_-}}\frac{\xi\left(\langle\lambda|\alpha\rangle\right)}{\xi \left(1+\langle\lambda|\alpha\rangle\right)}
=\prod_{\substack{\alpha \in \Delta_+\\w\alpha \in \Delta_-}} c\left(\langle \lambda|\alpha\rangle\right)\,.
\end{align}
The product runs over all positive roots, which also satisfy the condition that $w\alpha$ be a negative root for the Weyl group element $w$. The function $\xi$ is the completed Riemann $\zeta$-function and is defined as $\xi(k)\equiv\pi^{-k/2}\Gamma\left(\frac{k}{2}\right)\zeta(k)$.\footnote{The Riemann function can be seen to occur by using the $p$-adic approach to automorphic functions~\cite{Langlands,GarlLoop}.} The expansion (\ref{minparexp}) will be referred to as {\em minimal parabolic expansion} of the constant terms.

There is another way of arranging the constant terms that corresponds to choosing a maximal parabolic subgroups defined by a node $j_\circ$ as in (\ref{MaxParDef}). 
In order to introduce it, let us remark that the Levi component $M_{j_\circ}$ can be written as the product of two groups, namely
\begin{align}
\label{LeviMax}
M_{j_\circ}=GL\left(1\right)\times G_d\,,
\end{align}
where $G_d$ is the subgroup of $E_{d+1}$ which is determined by our choice of a simple root $\alpha_{j_\circ}$ in the Dynkin diagram of $E_{d+1}$. The Dynkin diagram of $G_d$ is given by the diagram which is left once one has deleted the node associated with $\alpha_{j_\circ}$ from the Dynkin diagram of $E_{d+1}$. The one-parameter group $GL\left(1\right)$ can be parameterised by a single variable $r\in \mathbb{R}^\times$.

The corresponding arrangement then highlights the dependence on only one of the parameters, namely $r$, corresponding to the single node $j_\circ$ (say, a decompactifying circle) and maintains the invariance under the remaining group $G_d$ in the decomposition (\ref{LeviMax}). In that case the constant terms can be packaged using cosets of the Weyl group $\mathcal{W}$. Denoting the Weyl group of $P_{j_\circ}$ by $\mathcal{W}_{P_{j_\circ}}$, the constant terms read~\cite{KS,GRS,Moeglin,GreenESeries} 
\begin{align}\label{parabexp}
\int\limits_{N_{P_{j_\circ}}(\mathbb{Z})\backslash N_{P_{j_\circ}}(\mathbb{R})} 
\!\!\!\!\!\!\!\! E^G(\lambda,g) \mathrm{d}n = \sum_{w\in \mathcal{W}_{j_\circ}\backslash\mathcal{W}} M(w,\lambda)e^{\langle\left(w\lambda+\rho\right)_{\parallel j_\circ}|H(g)\rangle}E^{G_d}\left(\left(w\lambda\right)_{\perp j_\circ},g\right)\,.
\end{align}
Let us explain some of the notation introduced here. For a weight $\lambda$, $(\lambda)_{\parallel j_\circ}$ is a projection operator on the component of $\lambda$ proportional to the fundamental weight $\Lambda_{j_\circ}$, and $(\lambda)_{\perp j_\circ}$ is orthogonal to $\Lambda_{j_\circ}$, i.e., a linear combination of the simple roots of $G_d$. The Eisenstein series on the right does not depend on the $GL(1)$ factor in (\ref{LeviMax}) since the dependence on the abelian group is explicitly factored out using the projections.
 The expression (\ref{parabexp}) does not depend solely on the Cartan subalgebra coordinates but also retains dependence on some of the positive step operators that appear in the Eisenstein series defined with respect to the reductive factor $G_d$. Even though indicated as depending on $g\in G$, the Eisenstein series on the r.h.s. of (\ref{parabexp}) effectively depends only on $g\in G_d$. This type of expansion is called {\em maximal parabolic expansion} of the constant terms of an Eisenstein series. 
 
For finite-dimensional groups the number of terms contained in the constant term in the minimal parabolic expansion (\ref{minparexp}) is generically equal to the finite order of the Weyl group and in the maximal parabolic expansion (\ref{parabexp}) to the number of Weyl group cosets.  For special choices of $\lambda$ there can be vast cancellations reducing the number of constant terms. These are the values that are relevant in string theory.
 
 For affine or general Kac--Moody groups one would expect generically infinitely many constant terms but again, as we will show, there are special choices for $\lambda$ where the number reduces to a small finite number. We will treat these cases in the next section but first describe more properties of the coefficients $M(w,\lambda)$ that control the cancellations.

\subsection{Functional relation and properties of $M(w,\lambda)$}

The factors $M(w,\lambda)$ are easily seen to satisfy the multiplicative identity
\begin{align}
\label{Mmult}
M(w\tilde{w},\lambda) = M(w,\tilde{w}(\lambda)) M(\tilde{w},\lambda)\,.
\end{align}
One also has the following functional relation for minimal Eisenstein series~\cite{Langlands}
\begin{align}
\label{funcrel}
E^G(\lambda,g) = M(w,\lambda) E^G(w(\lambda), g)\,.
\end{align}
This relation together with (\ref{Mmult}) is useful in showing that the sum in (\ref{parabexp}) is independent of the Weyl coset representative.

The completed Riemann function $\xi(s)$ entering in (\ref{LangCoeffs}) satisfies the simple functional equation
\begin{align}
\label{xifunc}
\xi(k) = \xi(1-k)\,,
\end{align}
which is at the heart of the meromorphic continuation of the Riemann $\zeta$-function.
Defining the function $c(k)$ by
\begin{align}
c(k):=\frac{\xi(k)}{\xi(1+k)}\,,
\end{align}
the functional equation (\ref{xifunc}) implies
\begin{align}\label{negC}
c(k)c(-k)=1.
\end{align}
The only (simple) zero of $c(k)$ occurs for $k=-1$; consequently $c(k)$ has a (simple) pole at $k=+1$:
\begin{align}
c(-1) = 0\,,\quad\quad c(+1) = \infty\,.
\end{align}
If, for a given $w$, the product $M(w,\lambda)$ contains more $c(-1)$ than $c(+1)$ factors, then $M(w,\lambda)$ will vanish.  This is exactly what happens for minimal Eisenstein series $E^G(\lambda,g)$ when $\lambda$ is chosen suitably as we will now explain in more detail (see also~\cite{GRS,Pioline:2010kb,GreenESeries,GreenSmallRep}).

We now restrict to the case of interest, namely $\lambda=2s\Lambda_{i_*}-\rho$, relevant for  the maximal parabolic Eisenstein series (\ref{maxparabfinite}). The argument of the $c$-function appearing in $M(w,\lambda)$ is $k=\langle\lambda|\alpha\rangle$. Now, for a simple root $\alpha_i\neq\alpha_{i_*}$
\begin{align}
k=\langle2s\Lambda_{i_*}-\rho|\alpha_i\rangle=-1\,.
\end{align}
Therefore, $c(\langle \lambda| \alpha_i\rangle)=0$ for simple roots $\alpha_i\neq \alpha_{i_*}$. This reduces the number of terms in the constant term considerably, namely one can restrict the sum over the Weyl group to the following subset~\cite{GreenESeries}
\begin{align}
\label{Si}
\mathcal{S}_{i_*}\equiv\left\{w\in\mathcal{W}|w\alpha_i>0\text{  for all  } i\neq i_*\right\}\subset \mathcal{W}\,.
\end{align}
If $w\notin\mathcal{S}_{i_*}$ then there will be at least one simple root $\alpha_i$ included in the product (\ref{LangCoeffs}) and consequently $M(w,\lambda)$ vanishes and the corresponding term in sum (\ref{minparexp}) disappears. The zero coming from the simple root cannot be cancelled by $c(+1)$ contributions from other roots; this can be argued by analytic continuation in $s$~\cite{GreenSmallRep}.

\subsection{Solving the condition in $\mathcal{S}_{i_*}$}

Now we want to give a more manageable description of the set $\mathcal{S}_{i_*}$ in (\ref{Si}). From the definition it follows that, as a set,
\begin{align}
\label{Squot}
\mathcal{S}_{i_*} = \mathcal{W} / \mathcal{W}_{i_*}\,,
\end{align}
where $\mathcal{W}_{i_*}$ is the Weyl group of the Levi factor $M_{i_*}$, i.e., the Weyl group described by the $E_{d+1}$ diagram where the node $i_*$ has been removed; $\mathcal{W}_{i_*}$ can also be defined as the stabiliser in $\mathcal{W}$ of the fundamental weight $\Lambda_{i_*}$. The quotient (\ref{Squot}) has to arise since any non-trivial element in $\mathcal{W}_{i_*}$ maps at least one of the simple roots of $M_{i_*}$ to a negative root. Therefore we should remove any $\mathcal{W}_{i_*}$ element that appears at the right end of a Weyl word. Once this is done the Weyl words appearing in $\mathcal{S}_{i_*}$ start with $w_{i_*}$ on the right and will never map any positive root of $M_{i_*}$ to a negative root.

A different and more explicit description of this fact can be given by constructively computing the set $\mathcal{S}_{i_*}$ by using the Weyl orbit $\mathcal{O}_{i_*}$ of the fundamental weight $\Lambda_{i_*}$. The Weyl words necessary for the orbit $\mathcal{O}_{i_*}$ are exactly those appearing in $\mathcal{S}_{i_*}$.

We illustrate the procedure for the specific example of ${E_8}$ and $i_*=1$, so that the Dynkin labels of the fundamental weight are  $\Lambda_1=[1,0,0,0,0,0,0,0]$. The only fundamental Weyl reflection that acts non-trivially on $\Lambda_1$ is $w_1$,  yielding the weight $[-1,0,1,0,0,0,0,0]$. In order to create a new weight we can only act with $w_3$, yielding $[0,0,-1,1,0,0,0,0]$. Then one can only act with $w_4$, giving $[0,1,0,-1,10,0,0]$. At this point we have two possibilities of fundamental Weyl reflections to act with, namely $w_2$ and $w_5$, giving us $[0,-1,0,0,1,0,0,0]$ and $[0,1,0,0,-1,1,0,0]$ respectively. We continue in this way iteratively until we are left with weights with entries being only $-1$ or $0$.\footnote{This only happens for finite-dimensional Weyl groups and the final element in the orbit is the negative of a dominant weight.} The first few Weyl words generated in this way are summarised in Table \ref{Lambda1Orb}. In this way one computes efficiently all the elements of $\mathcal{S}_{i_*}$ from the orbit of $\Lambda_{i_*}$.

\begin{table}
\begin{center}
\begin{tabular}{ | c | c |  }
  \hline                       
  Weyl words & Weights in Orbit \\ \hline
  $\id$ & $\Lambda_1=[1,0,0,0,0,0,0,0]$ (dominant weight) \\
  $w_{i_*}=w_1$ & $[-1,0,1,0,0,0,0,0]$ \\
  $w_3w_1$ & $[0,0,-1,1,0,0,0,0]$ \\
  $w_4w_3w_1$ & $[0,1,0,-1,1,0,0,0]$ \\
  $w_2w_4w_3w_1$; $w_5w_4w_3w_1$ & $[0,-1,0,0,1,0,0,0]$; $[0,1,0,0,-1,1,0,0]$ \\ 
  \vdots & \vdots \\ \hline 
\end{tabular}
\caption{\em Weyl words and weights in the Weyl orbit of $\Lambda_1$ for $E_8$.
\label{Lambda1Orb}}
\end{center}
\end{table}

The size $|\mathcal{O}_{i_*}|$ of the Weyl orbit of $\Lambda_{i_*}$ in the finite-dimensional case is given by
\begin{align}
|\mathcal{O}_{i_*}|=\frac{|\mathcal{W}|}{|\textrm{stab}(\Lambda_{i_*})|}=\frac{|\mathcal{W}|}{|\mathcal{W}_{i_*}|}\,.
\end{align}
For our example $\textrm{stab}(\Lambda_1)=\mathcal{W}(D_7)$ and the size of the orbit is $2160$. Therefore we have $2160$ distinct Weyl words in the left column of Table \ref{Lambda1Orb}.

We now prove formally that each Weyl word $w$ that generates an element of $\mathcal{O}_{i_*}$ satisfies $w\alpha_{i}>0$ for $i\neq i_*$. This establishes a one-to-one correspondence between elements in the Weyl orbit $\mathcal{O}_{i_*}$ and $\mathcal{S}_{i_*}$. The proof is by induction (on the length of the Weyl word/height of the weight in the orbit).

The identity element is in $\mathcal{S}_{i_*}$ and corresponds to the weight $\Lambda_{i_*}$. Suppose now that a particular Weyl word $w\in \mathcal{S}_{i_*}$ corresponds to a weight $w(\Lambda_{i_*})$ in the orbit $\mathcal{O}_{i_*}$. To continue the orbit we need to analyse the Dynkin labels of $w(\Lambda_{i_*})$; these are given by $p_i=\langle w(\Lambda_{i_*})| \alpha_i\rangle$ for $i=1,\ldots,\rk(G)$. We have to distinguish the three cases when a given $p_i$ is positive, negative or vanishes, and consider in all cases whether we $w_i w$ is in $\mathcal{S}_{i_*}$. 

First suppose that we have (for a fixed $i$)
\begin{align}
p_i = \langle w(\Lambda_{i_*})|\alpha_i\rangle=0\,,
\end{align}
where $\alpha_i$ is the $i^{\textrm{th}}$ simple root. By invariance of the product we also have $\langle \Lambda_{i_*}|w^{-1}(\alpha_i)\rangle=0$. From this we see that the root $w^{-1}(\alpha_i)=:\alpha_M$ is a linear combination of all simple roots other than $\alpha_{i_*}$. i.e., it is a root of the Levi factor $M_{i_*}$. Writing $\alpha_M=\sum_{k\neq i_*} n_k \alpha_k$ either all $n_k$ are non-negative or non-positive. Applying $w$ to $\alpha_M$ yields $\alpha_i=w(\alpha_M)=\sum_{n\neq i_*} n_k w(\alpha_k)$. But by assumption $w(\alpha_k)>0$ for all $k\neq i_*$; the equation can only be true if $\alpha_M=\alpha_j$ for some $j\neq i_*$. But this implies  immediately that
\begin{align}
w_iw(\alpha_j)=w_i(\alpha_i)=-\alpha_i<0
\end{align}
and we conclude that $w_iw\notin\mathcal{S}_{i_*}$. Similarly, if $p_i=0$ then $w_i$ will leave $w(\Lambda_{i_*})$ invariant and therefore $w_iw(\Lambda_{i_*})$ does not produce a new element of the Weyl orbit.

Secondly, we consider the case of
\begin{align}
p_i= \langle w(\Lambda_{i_*})|\alpha_i\rangle>0\,.
\end{align}
By invariance again $\langle \Lambda_{i_*}|w^{-1}(\alpha_i)\rangle>0$ and we conclude that
\begin{align}
w^{-1}(\alpha_i)=p_i\alpha_{i_*}+\text{(positive linear combination of $\alpha_{i\neq i_*}$'s)}\,,
\end{align}
where $p_i\in\mathbb{Z}_{>0}$. Now suppose $w_iw(\alpha_j)<0$ for some $j\neq i_*$. This can only happen if $w(\alpha_j)=\alpha_i$ since $\alpha_i$ is the only positive root that is mapped to a negative root by $w_i$ and $w(\alpha_j)$ is positive by the induction assumption. But then $\alpha_j = w^{-1}(\alpha_i)$ which cannot happen since $w^{-1}(\alpha_i)$ has a non-vanishing component along $\alpha_{i_*}$. Therefore $w_iw(\alpha_j)>0$ for all $j\neq i_*$ and therefore $w_iw\in S_{i_*}$. Similarly, when $p_i>0$ the element $w_iw(\Lambda_{i_*})$ has a lower height than $w(\Lambda_{i_*})$ and hence is also a new element of the orbit $\mathcal{O}_{i_*}$.

Finally, we consider the case 
\begin{align}
p_i= \langle w(\Lambda_{i_*})|\alpha_i\rangle<0\,.
\end{align}
Here, $w_iw(\Lambda_{i_*})=w(\Lambda_{i_*})-p_i \alpha_i$ and hence the height $w_iw(\Lambda_{i_*})$ is larger than that of $w(\Lambda_{i_*})$ and is an element of the orbit that has already computed. But this means that $w_iw$ has an equivalent representative in $\mathcal{W}$ of shorter length that has already been accounted for in $\mathcal{S}_{i_*}$. Therefore, the element $w_iw$ is in $\mathcal{S}_{i_*}$ but not a new one in the same way that $w_iw(\Lambda_{i_*})$ is not a new element of the Weyl orbit $\mathcal{O}_{i_*}$. This completes the proof.

In summary,  there is a one-to-one correspondence between the elements of $\mathcal{S}_{i_*}$ and Weyl words that make up the orbit $\mathcal{O}_{i_*}$. This correspondence gives also a very manageable way of constructing the set $\mathcal{S}_{i_*}$ by starting from the dominant weight $\Lambda_{i_*}$ and computing its Weyl orbit as a rooted and branched tree of Weyl words of increasing length.\footnote{There is a natural partial order induced on the constant terms from the Weyl orbit; this can be used to display the constant term structure in terms of a Hasse diagram.} By the multiplicative identity (\ref{Mmult}), one obtains that when going down the tree one has that if $M(\tilde{w},\lambda)$ vanishes, the subsequent $M(w\tilde{w},\lambda)$ will also vanish. Therefore one can stop the construction of the tree along a given branch once the factor $M(w,\lambda)$ on a vertex vanishes.\footnote{Again, it cannot happen that the zero of $M(\tilde{w},\lambda)$ gets balanced by a diverging $M(w,\tilde{w}(\lambda))$.}

The analysis in this section can clearly be extended to the case where $\lambda+\rho$ entering in the definition (\ref{minimalparabfinite}) of the minimal parabolic Eisenstein series is not proportional to a single $\Lambda_{i_*}$ but has support on several fundamental weights. The contributing Weyl words are still in one-to-one correspondence with the orbit of $\lambda+\rho$.

The restriction of the sum to the quotient $\mathcal{W}/\mathcal{W}_{i_*}$ for the constant terms expanded in the minimal parabolic subalgebra has also consequences for the expansion in maximal parabolic algebras as described by  formula (\ref{parabexp}). The constant terms in this case are described by {\em double cosets} via (see also~\cite{GRS})
\begin{align}
\label{ConstMaxFin}
&\int\limits_{N_{P_{j_\circ}}(\mathbb{Z})\backslash N_{P_{j_\circ}}(\mathbb{R})} 
E^G(\lambda,g) \mathrm{d}n \nn\\
&\quad = \sum_{w\in \mathcal{W}_{j_\circ}\backslash\mathcal{W}/\mathcal{W}_{i_*}} M(w,\lambda)e^{\langle\left(w\lambda+\rho\right)_{\parallel j_\circ}|H(g)\rangle}E^{G_d}\left(\left(w\lambda\right)_{\perp j_\circ},g\right)\,.
\end{align}
These are typically very few in number. The rooted tree mentioned above can be contracted further in this case thanks to the double coset structure.

\subsection{The order $s$ and `guessing' the right Eisenstein series}

{}From the previous section we know that the constant term is given by a polynomial in the Cartan subalgebra coordinates with a total of at most $|\mathcal{O}_{i_*}|$ terms. This is the correct number of constant terms for generic $s$ but one can make the observation that for specific choices of the parameter $s$ only a small fraction of these terms will survive, with all the other terms being zero. The reason is that for such special choices of $s$, the factor $M(w,\lambda)$ (which of course depends on $s$ through $\lambda=2s\Lambda_{i_*}-\rho$) will vanish. This has the remarkable effect that even for large groups $G$, the number of constant terms is reduced drastically. The inner product $k=\langle \lambda|\alpha\rangle$ enters in $M(w,\lambda)$ via the factor $c(k)$ and the properties of the $c$-function imply that $M(w,\lambda)$ will only vanish if $k=-1$ for some $\alpha$. But this generically\footnote{There can be exceptions when $\alpha=n_{i_*}\alpha_{i_*}+\ldots$ for $n_{i_*}>1$. This arose in none of the cases we have considered. It is not fully inconceivable that for infinite-dimensional algebras such exceptions might happen since there $n_{i_*}$ is not bounded.} only happens if $\lambda$ is on the weight lattice and hence $2s\in \mathbb{Z}$; therefore the weight lattice plays a distinguished role.

{}As a mathematical exercise there could now be many interesting integral weights $\lambda$ to consider, maybe associated with general minimal parabolic Eisenstein series, but string theory suggests which $\lambda$ to select. As the Eisenstein series are meant to occur at a fixed order in the $\alpha'=\ell_s^{2}$ and T-duality $SO(d,d;\mathbb{Z})$ is an exact symmetry at each order in $\alpha'$~\cite{Narain:1985jj,Sen:1991zi,Meissner:1991ge}. That in particular the tree-level term --that we associate with the identity Weyl element in the expansion-- be invariant under T-duality implies that the weight $\lambda+\rho$ entering in the definition of the minimal parabolic Eisenstein series should be invariant under $SO(d,d;\mathbb{Z})\subset E_{d+1}(\mathbb{Z})$. In other words, $\lambda+\rho$ is proportional to $\Lambda_1$ (in the numbering of Fig.~\ref{fig:Edplus1Diag}), i.e., we immediately arrive at (for $d>3$)
\begin{align}
\lambda= 2s\Lambda_1 -\rho
\end{align}
for string theory applications of $E_{d+1}(\mathbb{Z})$ minimal parabolic Eisenstein series. This assumes that the whole function $\mathcal{E}_{(p,q)}^D$ is given by a (single) minimal parabolic Eisenstein series, something that is not true for all $p$ and $q$. Choosing a weight determined by $\Lambda_1$ is also the only way of getting string perturbation theory right, see (\ref{pertlim}) below. A similar conclusion was reached in~\cite{Gubay:2010nd}. One could use the functional relation (\ref{funcrel}) to replace $\lambda$ by any element in its Weyl orbit.

The only remaining question then is to fix the parameter $s$ for the various types of higher derivative corrections. This can be done for example as follows. Supposing one knows the Laplace eigenvalue of the Eisenstein series from different considerations (e.g., as in~\cite{Green:2010wi}), then one needs to fix $s$ such that the quadratic Casimir gives the correct value.\footnote{If one knew all eigenvalues under the full set of higher order $E_{d+1}$ invariant differential operators one could determine $\lambda$ without making recourse to T-duality invariance. Another comment is that it is not {\em a priori} clear that the value of $s$ is independent of the dimension. It turns out that this can be achieved for $R^4$ and $\partial^4R^4$.} For the $R^4$ curvature correction term at order $(\alpha')^3$ this implies $s=3/2$, and  for the $\partial^4R^4$ curvature correction term at order $(\alpha')^5$ this gives $s=5/2$. Alternatively, one can compute this from the leading wall in a cosmological billiard (BKL) analysis, see~\cite{CurvCorrDamour,CurvCorrHanany} and section~\ref{sec:Laplace} below. This would immediately give $s=n/2$ at order $(\alpha')^n$. Finally, $s$ can be determined from comparing to known results from string scattering calculations, e.g.~\cite{Gross:1986iv,Kiritsis:1997em} and (\ref{pertlim}) below.

By the functional relation (\ref{funcrel}) one can also check which terms lift to $D=11$; this requires that there is a Weyl-equivalent $\lambda'=w(\lambda)$ such that $\lambda'+\rho$ is integrally proportional to $\Lambda_2$. This happens for $s=3/2$ but not for $s=5/2$, consistent with the fact that there is a $R^4$ curvature correction term in $D=11$, whereas there is no such $\partial^4R^4$ term.

That the corresponding Eisenstein series for $5\geq D\geq 3$ and $D=10$, normalised as in (\ref{ESeries}) produce the right constant terms and abelian Fourier coefficients was checked in~\cite{Pioline:2010kb,GreenESeries,GreenSmallRep}. In dimensions $6\leq D \leq 9$ the coefficient functions $\mathcal{E}^D_{(0,0)}$ and $\mathcal{E}^D_{(1,0)}$ are also known as sums of Eisenstein series~\cite{GreenESeries}.

\section{Constant terms: infinite-dimensional case}
\label{sec:CT}

The constant term in the full expansion of the maximal parabolic Eisenstein series over an affine group is given by
\begin{align}
\label{FullConstTermAffine}
\int\limits_{\hat{N}(\mathbb{Z})\backslash \hat{N}(\mathbb{R})} E^G(\hat{\lambda};\hat{g},v) \mathrm{d}\hat{n} =\sum_{\hat{w}\in\widehat{\mathcal{W}}}
M(\hat{w},\hat{\lambda})
e^{\langle \hat{w}\hat{\lambda}+\hat{\rho}|\hat{H}(v^d\,\hat{g})\rangle}\,.
\end{align}
The constant term in the expansion with respect to a particular maximal parabolic subgroup $P_{j_\circ}$ is given by\footnote{Note that the Levi factor in this case is a finite-dimensional group.}
\begin{align}
\label{ParabConstTermAffine}
&\quad\quad\quad\quad \int\limits_{N_{P_{j_\circ}}(\mathbb{Z})\backslash N_{P_{j_\circ}}(\mathbb{R})}
 E^G(\hat{\lambda};\hat{g},v) \mathrm{d}n  \nonumber\\
 &=
\sum_{\hat{w}\in \mathcal{W}_{j_\circ}\backslash\widehat{\mathcal{W}}} M(\hat{w},\hat{\lambda})e^{\langle\left(\hat{w}\hat{\lambda}+\hat{\rho}\right)_{\parallel j_\circ}|\hat{H}(v^d\,\hat{g})\rangle}E^{G_d}\left(\left(\hat{w}\hat{\lambda}\right)_{\perp j_\circ},\hat{g}\right)\,.&
\end{align}
where in all the formul\ae~$\hat{\lambda}=2s\hat{\Lambda}_{i_*}-\hat{\rho}$, so that we are again restricting to maximal parabolic Eisenstein series. The projections $(\hat{\lambda})_{\parallel j_\circ}$ and $(\hat{\lambda})_{\perp j_\circ}$ are different now from those in (\ref{parabexp}) since the Cartan subalgebra includes the additional direction $d$. $(\hat{\lambda})_{\perp j_\circ}$ has to be a weight of the Levi factor $M_{j_\circ}$ and has two directions less than $\hat{\lambda}$; it is a combination of the simple roots of $G_d$. By contrast, $(\hat{\lambda})_{\parallel j_\circ}$ is a combination of the fundamental weight $\hat{\Lambda}_{j_\circ}$ and the null root $\hat{\delta}$. The Levi factor explicitly reads
\begin{align}
M_{j_\circ} = GL(1)\times GL(1) \times G_d
\end{align}
and the pre-factor of the Eisenstein series in (\ref{ParabConstTermAffine}) now depends on the two parameters of the $GL(1)$ factors. One of them is $v$ and we will call the other one $r$ below.

In the affine case the expressions above follow from~\cite{GarlLoop}. We will assume that they also hold {\em mutatis mutandis} in the general Kac--Moody case (where one does not need $v$ and they therefore are similar to (\ref{minparexp}) and (\ref{parabexp})) and provide some consistency checks on this assumption with our calculations.
The validity of~\eqref{FullConstTermAffine}, i.e. convergence of the series, is in proven in~\cite{GarlLoop} for the affine case. In particular it was proven that~\eqref{FullConstTermAffine} possesses a meromorphic continuation, which extends the convergence condition stated for equation~\eqref{ESeriesAffine} to $\textrm{Re}\langle \hat\lambda | \hat{\delta}\rangle>-\text{ht}(\hat{\delta})$. 

For the finite-dimensional groups we have seen that the number of terms in the constant term of the maximal parabolic Eisenstein series is bounded from above by the size of the Weyl orbit $|\mathcal{O}_{i_*}|$, where $\alpha_{i_*}$ is the simple root with respect to which the maximal parabolic subgroup of the Eisenstein series is defined. Since the Weyl orbits of finite-dimensional groups are always of finite size, the constant term contains a finite number of terms. For the infinite-dimensional affine groups, however, the size of the Weyl orbits is infinite. Hence from our analysis given above and from equation \eqref{FullConstTermAffine} one would expect the constant term to be made up of an infinite number of terms. We will now show that for special choices of $s$, the number of constant terms for affine and other Kac--Moody groups can be reduced to a finite number. This is analogous to simplifications of constant terms in the case of finite-dimensional groups where the reduced number of terms is associated with small automorphic representations and required by string theory arguments.

\subsection{`Finite number of constant terms'-property}

The only way to reduce from an infinite to a finite number of terms is if for all but a finite number of terms in \eqref{FullConstTermAffine}, the coefficients $M(\hat{w},\hat{\lambda})$ vanish. The coefficients $M(\hat{w},\hat{\lambda})$, given by (\ref{LangCoeffs}), will vanish as before if they include more $c(-1)$ than $c(+1)$ factors.

In order to exhibit that almost all $M(\hat{w},\hat{\lambda})$ vanish for special $\hat{\lambda}$ we need some more notation and results on the affine root system~\cite{Kac90}. Let $G$ be a simple, simply-laced and maximally split Lie group as before; let $r=\rk(G)$ and denote by  $\alpha_i$ ($i=1,\ldots,r$) a choice of simple roots. In this basis  the unique highest root of $G$ is written as 
\begin{align}
\theta=\sum_{i=1}^r \theta_i \alpha_i = (\theta_i,\theta_2,\ldots,\theta_r)\,.
\end{align}
The affine extension of the root system is obtained adding a simple root $\alpha_0$.  From now on roots carrying a hat will be associated with roots of the affine group $\hat{G}$ whereas roots without a hat belong to $G$. A general affine root is then of the form
\begin{align}
\hat{\alpha}=n_0\alpha_0+n_1\alpha_1+...+n_r\alpha_r=n_0\hat{\delta}+\vec{\Delta}\cdot\vec{A}\,,
\end{align}
where we have used the standard definition of the null root 
\begin{align}
\hat{\delta}=\alpha_0+\theta
\end{align}
and introduced some further shorthand notation for finite-dimensional part of the root. The quantity $n_0$ is called the affine level and the vector $\vec{\Delta}$ is given by
\begin{align}
\vec{\Delta}=(n_1-n_0\theta_1,n_2-n_0\theta_2,...,n_r-n_0\theta_r)
\end{align}
and corresponds to a root vector of $G$ or vanishes.\footnote{Vanishing $\vec{\Delta}$ corresponds to imaginary roots of the algebra; they can never contribute to constant terms and therefore we will assume $\vec{\Delta}\neq 0$ in the following.}

Consider the expression $\langle\hat{\lambda}|\hat{\alpha}\rangle$ that appears in (\ref{LangCoeffs}) for $\hat{\lambda}=2s\hat{\Lambda}_{i_*}-\hat{\rho}$ and the affine Weyl vector 
 $\hat{\rho}$
\begin{align}
\langle\hat{\lambda}|\hat{\alpha}\rangle = 2s\langle\hat{\Lambda}_{i_*}|\hat{\alpha}\rangle-\langle\hat{\rho}|\hat{\alpha}\rangle=2s\langle\hat{\Lambda}_{i_*}| \hat{\alpha}\rangle-\textrm{ht}(\hat{\alpha})\,,
\end{align}
with the height $\textrm{ht}(\hat{\alpha})=\sum_{i=0}^r n_i$.
We are interested in the condition $\langle\hat{\lambda}|\hat{\alpha}\rangle=\pm1$, where `$+$' corresponds to a $c(+1)$ factor and `$-$' to a $c(-1)$ factor in $M(\hat{w},\hat{\lambda})$. The condition $\langle\hat{\lambda}|\hat{\alpha}\rangle=\pm1$, together with the requirement that $\hat{\alpha}>0$ defines two sets of roots 
\begin{align}
\Delta_s(\pm1) := \left\{ \hat{\alpha} \;:\; \langle \hat{\lambda}|\hat{\alpha}\rangle 
=\langle 2s \hat{\Lambda}_{i_*}-\hat{\rho} | \hat{\alpha}\rangle =\pm 1\right\}\,.
\end{align}
Solving $\langle\hat{\lambda}|\hat{\alpha}\rangle=\pm1$  for $s$ we obtain
\begin{align}
\label{affineSeqn}
s=\frac{\textrm{ht}(\hat{\alpha})\pm1}{2\langle\hat{\Lambda}_{i_*}| \hat{\alpha}\rangle}=\frac{\textrm{ht}(\hat{\alpha})\pm1}{2n_{i_*}}\,.
\end{align}
We can express the height of $\hat{\alpha}$ as
\begin{align}
\textrm{ht}(\hat{\alpha}) = n_0\textrm{ht}(\hat{\delta})+\textrm{ht}(\vec{\Delta}\cdot\vec{A})=n_0\left(1+\sum_{i=1}^r\theta_i\right)+ \sum_{i=1}^r\Delta_i\,.
\end{align}
Further we note that when $i_*\neq0$, then $n_{i_*}=\Delta_{i_*}+n_0\theta_{i_*}$. Inserting both expressions into \eqref{affineSeqn} and solving for $n_0$ we obtain
\begin{align}
\label{n0toSrelation}
n_0=\frac{2s\Delta_{i_*}-\sum_{j=1}^r\Delta_j\mp1}{\textrm{ht}(\hat{\delta})-2s\theta_{i_*}}
\end{align}
For a particular choice of the parameter $s$ and simple root $\alpha_{i_*}$, we can use this formula to determine the affine levels $n_0$ on which roots producing $c(\pm1)$  factors can occur. Since $-\theta_i\leq\Delta_i\leq\theta_i$, we see from the formula that there exists a maximum value of $n_0$, such that no roots producing $c(-1)$ or $c(+1)$ factors can exist on higher affine levels.\footnote{We assume that the denominator does not vanish. This is true in all cases of interest later.} In other words, both sets $\Delta_s(1)$ and $\Delta_s(-1)$ only contain a finite number of elements. The result and formula (\ref{n0toSrelation}) remain true if $i_*=0$ and one declares $\theta_0=0$.

Having determined the roots which may possibly cause the coefficient factor $M(\hat{w},\hat{\alpha})$ to vanish we now determine for which $\hat{w}$ they actually contribute in the product running over positive roots.
A root $\hat{\alpha} \in \Delta_s(\pm 1)$ will only appear in the product defining $M(\hat{w},\hat{\lambda})$, if for a particular Weyl word $\hat{w}$, the condition $\hat{w}(\hat{\alpha})<0$ is satisfied. In order to analyse this condition, we need to consider the general action of an affine Weyl group element $\hat{w}$.

The Weyl group $\widehat{\mathcal{W}}$ of an affine algebra can be written as a semi-direct product of the classical Weyl group $\mathcal{W}$ and a translational part $\mathcal{T}\cong \mathbb{Z}^r$ (where $r$ is the rank of the underlying finite-dimensional algebra)
\begin{align}
\widehat{\mathcal{W}}=\mathcal{W}\ltimes\mathcal{T}\,.
\end{align}
We will write an element of $\widehat{\mathcal{W}}$ as $\hat{w}=(w,t_\beta)$, where $w\in\mathcal{W}$ and $t_\beta\in\mathcal{T}$ with $\beta$ an element of the finite-dimensional root lattice. It should be noted that in general $\beta$ is \textit{not} a root of the algebra. The action of $\hat{w}$ on a general root $\hat{\alpha}=n_0\hat{\delta}+\vec{\Delta}\cdot\vec{A}$ is then given by
\begin{align}
\label{walphageneral}
\hat{w}(\hat{\alpha})=(w,t_\beta)(\hat{\alpha})&=w\left(t_\beta(\hat{\alpha})\right)&\nn\\
&=w\left(\hat{\alpha}-\langle\vec{\Delta}\cdot\vec{A}|\beta\rangle\hat{\delta}\right)&\nn\\
&=w\left(\vec{\Delta}\cdot\vec{A}+(n_0-\langle\vec{\Delta}\cdot\vec{A}|\beta\rangle)\hat{\delta}\right)&\nn\\
&=w(\vec{\Delta}\cdot\vec{A})+\left(n_0-\sum_{i=1}^r\Delta_i\langle \alpha_i|\beta\rangle\right)\hat{\delta}\,.&
\end{align}
{}From the last line of \eqref{walphageneral}, we conclude that for a $\beta$ of sufficient height (corresponding to $\hat{w}$ of sufficient length) and appropriate direction, the coefficient of the null root $\hat{\delta}$ will be negative and therefore we have $\hat{w}(\hat{\alpha})<0$. Then the root $\hat{\alpha}$ will appear in the product expression for $M(\hat{w},\hat{\lambda})$ and will produce a $c(\pm 1)$-factor. The conditions on $\beta$ will always be satisfied for almost all $\hat{w}$ that contribute to the constant term. We now show this in an example.

\subsection{Example: $\widehat{SL(2,\mathbb{R})}$}

In the following we consider the maximal parabolic Eisenstein series $E^{\hat{A}_1}_{\alpha_{i_*};s}$ for the affine extension $\hat{A}_1$ of $A_1= SL(2,\mathbb{R})$. In this example we will choose $\alpha_{i_*}$ to be determined by $i_*=1$. The root system of $\hat{A}_1$ is given by
\begin{align}
\hat{\alpha}=n_0\alpha_0+n_1\alpha_1=n_0\hat{\delta}+\Delta_1\alpha_1\,,
\end{align}
with integers $n_0$ and $n_1$ such that $n_0-n_1\in\{-1,0,1\}$. Here, $\hat{\delta}=\alpha_0+\alpha_1$ and $n_0$ counts the affine level. The height is $\textrm{ht}(\hat{\alpha})=n_0+n_1$.

In order to gain some intuition let us briefly consider the affine Weyl group orbit $\mathcal{O}_{i_*}$. Starting with the fundamental weight $\Lambda_{i_*}$ we construct its Weyl orbit in a similar way to the one already described for the case of finite-dimensional groups. We obtain Table \ref{affineorbit}.

\begin{table}
\begin{center}
\begin{tabular}{ | c | c |  }
  \hline                       
  Weyl words & Weights in Orbit \\ \hline
  $\id$ & $\Lambda_{1}=[0,1]$ (dominant weight) \\
  $w_{i_*}=w_1$ & $[2,-1]$ \\
  $w_0w_1$ & $[-2,3]$ \\
  $w_1w_0w_1$ & $[4,-3]$ \\
  $w_0w_1w_0w_1$ & $[-4,5]$ \\ 
  \vdots & \vdots \\ \hline 
\end{tabular}
\caption{\em Affine Weyl orbit of $\Lambda_{i_*=1}$.
\label{affineorbit}}
\end{center}
\end{table}

It is easy to see that we obtain an infinite number of weights in this orbit. The Weyl words in the left column of the table make up the set $\mathcal{S}_{i_*=1}^\infty$ and satisfy the condition $\hat{w}(\alpha_0)>0$ for all $\hat{w}\in\mathcal{S}_1^\infty$. Here, we have added $\infty$ to indicate that $\mathcal{S}_1^\infty$ contains an infinite number of elements.

In the notation introduced above, the set of elements $\mathcal{S}_1^\infty$ is given by
\begin{align}\label{S1infty}
\mathcal{S}_1^\infty=\left\{(\id,t_{k\alpha_1})\right\}_{k\in\mathbb{Z}_{\geq0}}\cup\left\{(w_1,t_{k\alpha_1})\right\}_{k\in\mathbb{Z}_{\geq0}}\,,
\end{align}
where $t_{\alpha_1}=w_0w_1$. From equation \eqref{walphageneral} we see that the action of an element $\hat{w}\in\mathcal{S}_1^\infty$ becomes
\begin{align}
\hat{w}(\hat{\alpha})=w(\Delta_1\alpha_1)+(n_0-2\Delta_1 k)\hat{\delta}\,.
\end{align}
{}From the second term in this equation we conclude that $\hat{w}(\hat{\alpha})$ will be a negative root for $\Delta_1=1$ and long enough Weyl words $\hat{w}$ (large enough $k$). For $\Delta_1=1$ we see from \eqref{affineSeqn} that we will get $c(-1)=0$ factors in $M(\hat{w},\hat{\lambda})$ for $s=(n_1-1)/n_1$ with $n_1\in \mathbb{Z}_{>0}$, i.e. $s=0,1/2,2/3,3/4,4/5,\ldots$. For these choices of $s$ the constant term will contain a finite number of terms since there are no cancellations from $c(+1)$ factors.

\subsection{$E_9$, $E_{10}$ and beyond}

In the case of $E_9$ it is not so simple to write down the set $\mathcal{S}_{i_*}^\infty$ in an equally explicit way as was done for the case of $\hat{A}_1$ in \eqref{S1infty}. However, the argument we gave in \eqref{walphageneral}, that a root $\hat{\alpha}$ will become negative when acting on it with a long enough Weyl word from the set $\mathcal{S}_{i_*}^{\infty}$ still holds.  From relation \eqref{n0toSrelation} one can then see again that both sets $\Delta_s(\pm1)$ contain a finite number of roots. In practice, one can first compute the finite sets $\Delta_s(\pm 1)$ and then construct the set $\mathcal{S}_{i_*}^{\infty}$ iteratively from the Weyl orbit $\mathcal{O}_{i_*}$ and check whether after a finite number of steps it happens that more elements from $\Delta_s(-1)$ than from $\Delta_s(+1)$ contribute to $M(\hat{w},\hat{\lambda})$. By the multiplicative identity (\ref{Mmult}) one then can terminate the calculation of $\mathcal{S}_{i_*}$ along the branch of the orbit where this happened. If $\hat{\lambda}$ is chosen appropriately only a (small) finite number of Weyl words remain in $\mathcal{S}_{i_*}^{\infty}$ and give contributions to the constant terms.

Due to the absence of the nice affine level structure, the situation for hyperbolic Kac-Moody algebras is much harder to analyse. It is not possible to use a formula similar to \eqref{n0toSrelation}  to see that the sets $\Delta_s(\pm1)$ only contain a finite number of elements. Instead one can use the following procedure for  Eisenstein series with weight\footnote{Now, all the quantities refer to the hyperbolic algebra but we refrain from putting additional decorations on the symbols to avoid cluttering the notation.} $\lambda=2s\Lambda_{i_*}-\rho$. The relevant inner product is 
\begin{align}
\langle \lambda | \alpha\rangle= 2s n_{i_*} - \textrm{ht}(\alpha)\,.
\end{align}
The height of a root grows much faster than the component along a given root $n_{i_*}$. It is hence clear that for moderately small $s$  roots of sufficient height will have inner products $\langle \lambda | \alpha\rangle<-1$ and therefore will not belong to $\Delta_s(\pm 1)$. Therefore, computing the set of `dangerous' roots $\Delta_s(\pm 1)$ is a finite computational problem. More precisely, we can denote by $\Delta(n_{i_*})$ the set of positive real roots $\alpha=\sum_i n_i \alpha_i$ with a given $n_{i_*}$. This set is finite as long as the removal of the node $i_*$ from the Dynkin diagram leaves the diagram of a finite-dimensional algebra. For $E_{10}$ this means $i_*\neq 10$. We will assume this in the following. Then we can define
\begin{align}
h(n_{i_*}):= \textrm{min}\left\{\textrm{ht}(\alpha)\;:\; \alpha\in \Delta(n_{i_*})\right\}\,.
\end{align}
This is a monotonous function of $n_{i_*}$. Its rate of growth with $n_{i_*}$ is roughly equal to the height of the affine null root of the underlying affine algebra divided by its Kac label. For moderately small $s$ --~like those of interest to us~-- this is greater than the rate of growth of $2sn_{i_*}$. Therefore we can construct $\Delta(n_{i_*})$ by increasing height and terminate the construction of roots when $2s n_{i_*}-h(n_{i_*})<-1$ for some $n_{i_*}$.\footnote{To be on the safe side computationally, one can check the next few steps after this inequality is satisfied for the first time.} From the resulting finite set of roots we can select those $\alpha$ that belong to $\Delta_s(\pm 1)$. 

The next step is to determine those Weyl words that contribute to the constant terms. This is done in the same way as before: One constructs the Weyl words from the orbit of $\Lambda_{i_*}$ and checks whether more elements from $\Delta_s(-1)$ than from $\Delta_s(+1)$ contribute to $M(w,\lambda)$. For generic $s$ this will of course result in an infinite number of Weyl words. However, if $s$ is chosen appropriately, this leaves a finite number of Weyl words and hence a finite number of summands in the constant term. These are the cases that we will focus on in the following.

\section{Constant terms of $E_n(\mathbb{Z})$ Eisenstein series}
\label{sec:computations}

In this section, we present generalities on the calculations of the constant terms for the $E_n(\mathbb{Z})$ Eisenstein series for $n\leq 11$. In sections \ref{sec:DegLimit} to \ref{sec:DegLimitD1} we discuss the three possible degeneration limits and give explicit expressions for the respective maximal parabolic expansions. Minimal parabolic expansions are discussed in section \ref{sec:minimalexpansion} and the explicit expressions are given in appendix~\ref{sec:expansion}.

\subsection{Degeneration limits for $D\geq 3$}
\label{sec:DegLimit}

As mentioned in the introduction, important consistency checks of the functions $\mathcal{E}^D_{(p,q)}(\Phi)$ appearing in the analytic part of the four graviton scattering amplitude (\ref{FourGravAn}) in $D$ space-time dimensions\footnote{We do not discuss the issue of infrared divergences of these amplitudes in $D\leq 4$ here, nor their precise definition for $D\leq 3$.} are obtained by considering different degeneration limits of $\mathcal{E}^D_{(p,q)}$ in different corners of moduli space. The three limits  are referred to as the decompactification, perturbative and the semi-classical M-theory limit; and we restrict ourselves to taking the limit for the constant terms. What `taking the limit' means is to calculate the constant term of an Eisenstein series with respect to a particular maximal parabolic subgroup $P_{j_\circ}$. Formally, this corresponds to integrating out all the components of the unipotent radical $N_{j_\circ}$ of $P_{j_\circ}$ as in (\ref{parabexp}) and (\ref{ParabConstTermAffine}). We will use the following abbreviated notation for this integration
\begin{align}
\label{maxshort}
\int_{j_\circ}\mathcal{E}_{(p,q)}^D  \equiv  \int_{N_{j_\circ}/G(\mathbb{Z})\cap N_{j_\circ}}\,\mathcal{E}_{(p,q)}^D  \mathrm{d} n\,.
\end{align}

For $D\geq 3$, the parameter $r$ of the $GL(1)$ factor in the decomposition~\eqref{LeviMax}, acquires a different physical meaning in each of the three degeneration limits, and can be expressed in terms of fundamental string theory quantities. In \cite{GreenAutoProps,Pioline:2010kb,GreenESeries} general expressions for the three degeneration limits of $\mathcal{E}^D_{(0,0)}$ and $\mathcal{E}^D_{(1,0)}$ were given for $D\geq 3$, which we summarise for the readers' convenience.

\subsubsection*{Decompactification limit:}

In this limit $r_d/\ell_{D+1}\gg1$, which corresponds to making one of the circles of the torus very large in units of the $(D+1)$-dimensional Planck scale. In terms of maximal parabolic subgroups this limit corresponds to singling out the node $d+1$ in figure~\ref{fig:Edplus1Diag}, i.e., $j_\circ=d+1$, leading to $G_d = E_d$. One has the standard relation between Planck scales $\ell_{D+1}^{D-1}=\ell_D^{D-2}r_d$.  The constant terms of the coefficients $\mathcal{E}^{D}_{(0,0)}$ and $\mathcal{E}^{D}_{(1,0)}$ behave in the following way under expansion with respect to the parabolic subgroup $P_{\alpha_{d+1}}$~\cite{GreenAutoProps,Pioline:2010kb,GreenESeries} 
\begin{align}
\label{decompR41}
\int_{d+1}\mathcal{E}^{D}_{(0,0)}\simeq\frac{\ell^{8-D}_{D+1}}{\ell^{8-D}_{D}}\left(\frac{r_d}{\ell_{D+1}}\mathcal{E}^{D+1}_{(0,0)}+\left(\frac{r_d}{\ell_{D+1}}\right)^{8-D}\right)
\end{align}
and
\begin{align} 
\label{decompR42}
\int_{d+1}\mathcal{E}^{D}_{(1,0)}\simeq\frac{\ell^{12-D}_{D+1}}{\ell^{12-D}_{D}}\left(\frac{r_d}{\ell_{D+1}}\mathcal{E}^{D+1}_{(1,0)}+\left(\frac{r_d}{\ell_{D+1}}\right)^{6-D}\mathcal{E}^{D+1}_{(0,0)}+\left(\frac{r_d}{\ell_{D+1}}\right)^{12-D}\right)\,,
\end{align}
where the $\simeq$ symbol indicates that numerical factors in front of each term are not shown explicitly. The first terms on the right hand sides of the equations \eqref{decompR41} and \eqref{decompR42} are easily understood from decompactification from $D$ to $D+1$ dimensions; the other terms are threshold effects~\cite{GreenAutoProps}. Since one can relate $\ell_{D+1}/\ell_D$ to $r_d/\ell_{D+1}$, the expansion on the right hand side is in terms of a single variable that parameterises the $GL(1)$ in the Levi factor $M_{d+1} = GL(1)\times E_d$. In our conventions we have $r=(r_d/\ell_{D+1})^{(D-1)/(D-2)}=r_d/\ell_D$. This yields the following decompactification rules
\begin{align}
\label{declim}
\int_{d+1}\mathcal{E}^{D}_{(0,0)}&\simeq r^{6/(D-1)}\mathcal{E}^{D+1}_{(0,0)}+r^{8-D}\,,&\nn\\
\int_{d+1}\mathcal{E}^{D}_{(1,0)}&\simeq r^{10/(D-1)} \mathcal{E}^{D+1}_{(1,0)}+r^{D(D-7)/(1-D)}\mathcal{E}^{D+1}_{(0,0)}+r^{12-D}\,.&
\end{align}
These have to be fulfilled by the automorphic forms for $D\geq 3$. The coefficients of the last terms, that we call pure threshold terms, are known to be proportional to $\xi(8-D)$ and $\xi(12-D)$ respectively~\cite{GreenAutoProps,Pioline:2010kb}.

\subsubsection*{Perturbative limit:}

This corresponds to the weak string coupling expansion in $D$ dimensions $y_D\to 0$. The $D$-dimensional string coupling $y_D$ is given by $y_D=\ell^{D-2}_D/\ell_s^{D-2}$ and the string scale $\ell_s$ is kept fixed. Then one requires~\cite{GreenAutoProps,Pioline:2010kb,GreenESeries}
\begin{align}
\label{pertlim}
\int_{1}\mathcal{E}^{D}_{(0,0)}&\simeq\frac{\ell_s^{8-D}}{\ell_D^{8-D}}\left(\frac{2\zeta(3)}{y_D}+E^{SO(d,d)}_{d+1;\frac{d}{2}-1}\right)
\end{align}
and
\begin{align}
\int_{1}\mathcal{E}^{D}_{(1,0)}\simeq\frac{\ell_s^{12-D}}{\ell_D^{12-D}}\left(\frac{\zeta(5)}{y_D}+E^{SO(d,d)}_{d+1;\frac{d}{2}+1}+y_DE^{SO(d,d)}_{3;2}\right)\,,
\end{align}
respectively. Here, the first terms are fixed by string tree level calculations and the $SO(d,d)$ Eisenstein series on the right-hand side are maximal parabolic Eisenstein series as in (\ref{maxparabfinite}) and our (non-standard) labelling convention for the $SO(d,d)$ series is induced from removing node $1$ from the $E_{d+1}$ Dynkin diagram~\ref{fig:Edplus1Diag}. That is, the $d$ nodes are labelled $2$ through to $d+1$. Again one can recombine the pre-factors by using the definition of the string coupling and then expand in terms of a single variable which is associated to the $GL(1)$ factor in the Levi decomposition. We choose here $r=(\ell_s/\ell_D)^2=y_D^{2/(2-D)}$.  We note that the string coupling $y_D$ can be defined alternatively in terms of the ten-dimensional string coupling $g_s$ and the string compactification volume $V_d$ via $y_D= g_s^2 \ell_s^d/V_d$.

\subsubsection*{Semi-classical M-theory limit:}

In this limit one takes the volume of the whole M-theory torus large. In terms of the $E_{d+1}$ Dynkin diagram this corresponds to the maximal parabolic associated with node $2$. The relevant conditions on the Eisenstein series are then~\cite{GreenAutoProps,Pioline:2010kb,GreenESeries}
\begin{align}
\label{r4m1}
\int_{2}\mathcal{E}^{D}_{(0,0)}\simeq\frac{\mathcal{V}_{d+1}}{\ell^3_{11}\ell_D^{8-D}}\left(4\zeta(2)+\left(\frac{\ell_{11}^{d+1}}{\mathcal{V}_{d+1}}\right)^{\frac{3}{d+1}}E^{SL(d+1)}_{1;\frac32}\right)
\end{align}
and
\begin{align}
\label{r4m2}
\int_{2}\mathcal{E}^{D}_{(1,0)}&\simeq\frac{\ell_{11}\mathcal{V}_{d+1}}{\ell_D^{12-D}}\Bigg(\left(\frac{\mathcal{V}_{d+1}}{\ell_{11}^{d+1}}\right)^{\frac{1}{d+1}}E^{SL(d+1)}_{1;-\frac12}+\left(\frac{\ell_{11}^{d+1}}{\mathcal{V}_{d+1}}\right)^{\frac{5}{d+1}}E^{SL(d+1)}_{1;\frac52}\nn\\
&\quad+\left(\frac{\ell_{11}^{d+1}}{\mathcal{V}_{d+1}}\right)^{\frac{8}{d+1}}E^{SL(d+1)}_{3;2}\Bigg)\,.
\end{align}
The first term in (\ref{r4m1}) for the $R^4$ term is determined by a one-loop computation in $D=11$ supergravity~\cite{Green:1997as}, there is no similar term for $\partial^4R^4$ in (\ref{r4m2}) since this term does not exist as a curvature correction term in $D=11$. 

The parameter $r$ of the $GL(1)$ in the Levi factor of the maximal parabolic defined by node $2$ of the $E_{d+1}$ Dynkin diagram is then given by either
$r=(\mathcal{V}_{d+1}/\ell_{D}^{d+1})^{1/3}=(\mathcal{V}_{d+1}/\ell_{11}^{d+1})^{3/(D-2)}$, where $\ell_D^{D-2}=\ell_{11}^9/\mathcal{V}_{d+1}$, or equivalently $r^2= \mathcal{V}_{d+1}/\ell_{11}^3\ell_D^{8-D}$. Here, $\mathcal{V}_{d+1}$ denotes the volume of the M-theory torus (in contrast to the string theory torus $V_d$).

\subsection{Degeneration limits for $D=2$}
\label{sec:DegLimitD2}

When $D<3$ the limits above require additional care. This is due to the absence of a natural Planck length in $D=2$ space-time dimensions as normally defined through the two-derivative Einstein--Hilbert action; nor is it possible to define a Kaluza--Klein reduction from $D=3$ to $D=2$ such that one ends up in $D=2$ Einstein frame since the gravitational action is conformally invariant. Higher derivative terms on the other hand are of course accompanied by length scales.

\subsubsection*{Decompactification limit:}

In order to understand the decompactification limit from $D=3$ to $D=2$ one has to properly understand the relation between three-dimensional and two-dimensional gravity theories. This has been well-studied for example in the context of the Geroch group that describes the infinite symmetries of $D=2$ (super-)gravity (such as $E_9$). The set-up was  pioneered in~\cite{Geroch:1970nt,Julia:1982gx,Breitenlohner:1986um,Nicolai:1987kz} and reviewed for example in~\cite{Nicolai:1991tt,Maison:2000fj}.

The three-dimensional metric decomposes as (setting to zero the off-diagonal pieces for simplicity)
\begin{align}
\label{met2d}
ds_3^2 = \lambda^{-2} ds_2^2 + \rho^2 \left(dx^3\right)^2\,.
\end{align}
Here, $\lambda^{-1}$ is the conformal factor of the two-dimensional metric and $x^3$ is the compactifying direction. It is not possible to choose $\lambda$ such that the $D=2$ theory is in Einstein frame. One necessarily obtains {\em two} new parameters just like going from $E_8$ to $E_9$ enlarges the Cartan subalgebra by two generators.\footnote{In the context of the Geroch group, $\lambda$ is associated with the central extension and $\rho$ with the derivation~\cite{Julia:1980gr,Breitenlohner:1986um,Kleinschmidt:2005bq}. The same is true here.} The two parameters in (\ref{met2d}) are given by
\begin{align}
\label{2drels}
\lambda = \frac{\ell_3}{\ell_2}\,,\quad \rho = \frac{r_d}{\ell_3}\,,
\end{align}
where $r_d$ is the size of the decompactifying circle and we will refer to $\ell_2$ as the two-dimensional Planck scale. The two-derivative Einstein--Hilbert term in $D=2$ is not accompanied by the (arbitrary) length scale $\ell_2$, but the higher derivative terms are. The decompactification limit now consists in sending $\rho\to\infty$ and we choose to keep $\lambda$ fixed.

Performing the usual analysis of higher derivative couplings we obtain for the Eisenstein series the decompactification relations
\begin{align}
\label{2ddec}
\int_{d+1} \mathcal{E}^2_{(0,0)} &\simeq \lambda^{6} \left( \rho\mathcal{E}^3_{(0,0)} + \rho^{6} \right)\,,&\nn\\
\int_{d+1} \mathcal{E}^2_{(1,0)} &\simeq \lambda^{10} \left( \rho \mathcal{E}^3_{(1,0)} + \rho^{4} \mathcal{E}_{(0,0)}^3 + \rho^{10}\right)\,,&
\end{align}
where we have again suppressed numerical coefficients and augmented them by threshold terms as in (\ref{decompR41}). The decompactifying node is $d+1=3$ and unlike in other dimensions it is not possible to relate $\lambda$ and $\rho$. The precise numerical coefficients can be found in the detailed expansions of the Eisenstein series below where we will also see that requirement (\ref{2ddec}) forces us to modify the naive guess for the $D=2$ Eisenstein series.

\subsubsection*{Perturbative limit:}

The definition of the string coupling as above (\ref{pertlim}) fails in $D=2$, instead one should use the one at the end of that paragraph, i.e. $y_2= g_s^2 \ell_s^8/V_8$. Similar to the decompactification limit there is no way of relating the two-dimensional string coupling $y_2$ to the two-dimensional Planck length $\ell_2$, both appear as independent parameters. The perturbation limit on the automorphic form in terms of the $SO(d,d)$ invariant parameters $y_2$ and $\ell_s/\ell_2$ is then
\begin{align}
\label{2dpert}
\int_{1}\mathcal{E}^{2}_{(0,0)}&\simeq\left(\frac{\ell_s}{\ell_D}\right)^6\left(\frac{2\zeta(3)}{y_2}+E^{SO(8,8)}_{9;3}\right)\,,&\nn\\
\int_{1}\mathcal{E}^{2}_{(1,0)}&\simeq\left(\frac{\ell_s}{\ell_D}\right)^{10}\left(\frac{\zeta(5)}{y_2}+E^{SO(8,8)}_{9;5}+y_2E^{SO(8,8)}_{3;2}\right)\,.
\end{align}

\subsubsection*{Semi-classical M-theory limit:}

The relations (\ref{r4m1}) and (\ref{r4m2}) remain valid except that it is again impossible to relate the two-dimensional Planck length $\ell_2$ to the other variables and there are two independent $SL(9,\mathbb{Z})$ invariant expansion parameters, namely $\ell_2/\ell_{11}$ and the volume of the M-theory $9$-torus $\mathcal{V}_9/\ell_{11}^9$:
\begin{align}
\label{2dm}
\int_{2}\mathcal{E}^{2}_{(0,0)}\simeq\left(\frac{\ell_{11}}{\ell_2}\right)^6\left(4\zeta(2)\frac{\mathcal{V}_{9}}{\ell_{11}^{9}}+\left(\frac{\mathcal{V}_{9}}{\ell_{11}^{9}}\right)^{\frac{2}{3}}E^{SL(9)}_{1;\frac32}\right)
\end{align}
and
\begin{align}
\int_{2}\mathcal{E}^{2}_{(1,0)}&\simeq\left(\frac{\ell_{11}}{\ell_2}\right)^{10}\Bigg(\left(\frac{\mathcal{V}_{9}}{\ell_{11}^{9}}\right)^{\frac{10}{9}}E^{SL(9)}_{1;-\frac12}+\left(\frac{\mathcal{V}_{9}}{\ell_{11}^{9}}\right)^{\frac{4}{9}}E^{SL(9)}_{1;\frac52}+\left(\frac{\mathcal{V}_{9}}{\ell_{11}^{9}}\right)^{\frac{1}{9}}E^{SL(9)}_{3;2}\Bigg)\,.
\end{align}

\subsection{Degeneration limits for $D=1$}
\label{sec:DegLimitD1}

Since the dimension of the Cartan subalgebra of $E_{10}$  is equal to the number of simple roots of the algebra most limits are easier to describe than in the $E_9$ case. 

\subsubsection*{(Double) decompactification limit:}

The first limit we study is the decompactification limit which is the only problematic case since it involves two-dimensional gravity and the associated problems of conformal invariance. Equivalently, the maximal parabolic is the affine $E_9$.\footnote{For the algebraic relation between $E_9$ and $E_{10}$ see also~\cite{KMW}.} More precisely, it is again impossible to relate the ratio $r_d/\ell_2$ to the ratio of Planck scales $\ell_1/\ell_2$ since the two-dimensional Planck scale is ill-defined. But we note that the (pure) threshold terms in (\ref{decompR41}) and (\ref{decompR42}) are well-defined here since $\ell_2$ drops out. We did not determine from first principles the decompactification limit from $D=1$ to $D=2$ but instead a direct decompactification of two directions from $D=1$ to $D=3$. The general rule for this double decompactification (as implied for instance by (\ref{decompR41})) is
\begin{align}
\label{doubledec}
\int_{d+1,d} \mathcal{E}_{(0,0)}^D &\simeq v_2^6 \mathcal{E}_{(0,0)}^{D+2} + v_2^{D(7-D)} r^{D-6} + r^{8-D}\,,&
\end{align}
\begin{align}
\int_{d+1,d} \mathcal{E}_{(1,0)}^D &\simeq v_2^{10} \mathcal{E}_{(1,0)}^{D+2} + v_2^{(D+1)(6-D)} r^{D-4} \mathcal{E}_{(0,0)}^{D+2} +v_2^{D(11-D)} r^{D-10}&\nn\\
&\quad + v_2^6 r^{6-D} \mathcal{E}_{(0,0)}^{D+2} + v_2^{D(7-D)} 
+ r^{12-D}\,,&
\end{align}
where the expansion parameters are given in terms of the 2-torus volume 
\begin{align}
v_2 =\left(\frac{\text{vol}(T^2)\ell_{D+2}^{6-D}}{\ell_D^{8-D}}\right)^{1/6}
=\left(\frac{\text{vol}(T^2)\ell_{D+2}^{10-D}}{\ell_D^{12-D}}\right)^{1/10}
\end{align}
and one of the circles with $r= r_d/\ell_D$ as before. In the case $D=1$, these relations do not make explicit reference to the Planck length in two dimensions and remain well-defined. We will use the relation (\ref{doubledec}) to check our proposal for the $E_{10}(\mathbb{Z})$ Eisenstein series. Relating the $E_{10}(\mathbb{Z})$ series to $E_9(\mathbb{Z})$ we will also derive a single decompactification rule for $D=1$ that will turn out to be subtly different from (\ref{decompR42}) in the twelve derivative case. A double decompactification corresponds to a parabolic subgroup that is not maximal.

\subsubsection*{Perturbative limit:}

In this limit, the maximal parabolic subgroup has as semi-simple part the finite-dimensional $D_9=SO(9,9)$ T-duality group. The definitions of the expansion parameters in the cases $D>3$ continue to hold so that we immediately deduce
\begin{align}
\label{1dpert}
\int_{1}\mathcal{E}^{1}_{(0,0)}&\simeq\frac{\ell_s^{7}}{\ell_1^{7}}\left(\frac{2\zeta(3)}{y_1}+E^{SO(9,9)}_{10;\frac{7}{2}}\right)&\nn\\
&\simeq 2\zeta(3) y_1^6 +  y_1^7E^{SO(9,9)}_{10;\frac{7}{2}} \,,&
\end{align}
and
\begin{align}
\int_{1}\mathcal{E}^{1}_{(1,0)}&\simeq\frac{\ell_s^{11}}{\ell_1^{11}}\left(\frac{\zeta(5)}{y_1}+E^{SO(9,9)}_{10;\frac{11}{2}}+y_1E^{SO(9,9)}_{3;2}\right)&\nn\\
&\simeq \zeta(5)y_1^{10}+y_1^{11}E^{SO(9,9)}_{10;\frac{11}{2}}+y_1^{12}E^{SO(9,9)}_{3;2}
\,,&
\end{align}
where $y_1=\ell_s/\ell_1$ was used. Our expansion parameter $r$ below is related to $y_1$ via $r=y_1^2$.

\subsubsection*{Semi-classical M-theory limit:}

The maximal parabolic has now semi-simple factor $A_9=SL(10)$. The expressions (\ref{r4m1}) and (\ref{r4m2}) are still valid and become
\begin{align}
\label{1dm}
\int_{2}\mathcal{E}^{1}_{(0,0)}&\simeq\frac{\mathcal{V}_{10}}{\ell^3_{11}\ell_1^{7}}\left(4\zeta(2)+\left(\frac{\ell_{11}^{10}}{\mathcal{V}_{10}}\right)^{\frac{3}{10}}E^{SL(10)}_{1;\frac32}\right)&\nn\\
&\simeq4\zeta(2)\left(\frac{\mathcal{V}_{10}}{\ell_1^{10}}\right)^{2/3}+\left(\frac{\mathcal{V}_{10}}{\ell_1^{10}}\right)^{7/10}E^{SL(10)}_{1;\frac32}&
\end{align}
and
\begin{align}
\int_{2}\mathcal{E}^{1}_{(1,0)}&\simeq\frac{\ell_{11}\mathcal{V}_{10}}{\ell_1^{11}}\Bigg(\left(\frac{\mathcal{V}_{10}}{\ell_{11}^{10}}\right)^{\frac{1}{10}}E^{SL(10)}_{1;-\frac12}+\left(\frac{\ell_{11}^{10}}{\mathcal{V}_{10}}\right)^{\frac{5}{10}}E^{SL(10)}_{1;\frac52}+\left(\frac{\ell_{11}^{10}}{\mathcal{V}_{10}}\right)^{\frac{8}{10}}E^{SL(10)}_{3;2}\Bigg)\nn\\
&\simeq\left(\frac{\mathcal{V}_{10}}{\ell_{1}^{10}}\right)^{11/10}E^{SL(10)}_{1;-\frac12}+\left(\frac{\mathcal{V}_{10}}{\ell_{1}^{10}}\right)^{7/6}E^{SL(10)}_{1;\frac52}+\left(\frac{\mathcal{V}_{10}}{\ell_{1}^{10}}\right)^{6/5}E^{SL(10)}_{3;2}\,.
\end{align}
Our expansion parameter $r$ below is related to the fundamental quantities via $r=(\mathcal{V}_{10}/\ell_1^{10})^{1/3}$.

\subsection{Eisenstein series in $D<3$}

We propose that the $E_9$, $E_{10}$ and $E_{11}$  Eisenstein series that are relevant for the $R^4$ and $\partial^4 R^4$ terms are given by
\begin{align}
\label{e9eis}
\mathcal{E}_{(0,0)}^2 &= 2\zeta(3) v E^{E_9}_{1;3/2} \quad\quad&&\text{(i.e., $\hat{\lambda} = 3\hat{\Lambda}_1 +\hat{\delta}-\hat{\rho}$)},\nn\\
\mathcal{E}_{(1,0)}^2 &= \zeta(5) v E^{E_9}_{1;5/2} \quad\quad&&\text{(i.e., $\hat{\lambda} = 5\hat{\Lambda}_1 +\hat{\delta}-\hat{\rho}$)},
\end{align}
for $E_9$, by
\begin{align}
\label{e10eis}
\mathcal{E}_{(0,0)}^1 &= 2\zeta(3)  E^{E_{10}}_{1;3/2} \quad\quad&&\text{(i.e., $\lambda = 3\Lambda_1 -\rho$)},\nn\\
\mathcal{E}_{(1,0)}^1 &= \zeta(5)  E^{E_{10}}_{1;5/2} \quad\quad&&\text{(i.e., $\lambda = 5\Lambda_1 -\rho$)},
\end{align}
for $E_{10}$ and by
\begin{align}
\label{e11eis}
\mathcal{E}_{(0,0)}^0 &= 2\zeta(3)  E^{E_{11}}_{1;3/2} \quad\quad&&\text{(i.e., $\lambda = 3\Lambda_1 -\rho$)},\nn\\
\mathcal{E}_{(1,0)}^0 &= \zeta(5)  E^{E_{11}}_{1;5/2} \quad\quad&&\text{(i.e., $\lambda = 5\Lambda_1 -\rho$)},
\end{align}
for $E_{11}$. Except for the additional factor of $v$ related to the shift of the weight by $\hat{\delta}$ these are straight-forward generalisations of the results of~\cite{GreenAutoProps,Pioline:2010kb,GreenESeries}. (It is tempting to think that the addition of $\hat{\delta}$ means that the Eisenstein series is associated with (a lattice in) the so-called basic representation at level one~\cite{KMW}.)
In the following section we will subject the proposals for $E_9$ and $E_{10}$ to consistency checks by expanding the constant terms in different (maximal) parabolic subgroups and comparing to the degeneration limits discussed above. Our checks will only concern the constant terms and so are insensitive to possible cusp forms (which by definition have vanishing constant terms). In the finite-dimensional case, there are good arguments to show that no cusp forms compatible with string theory boundary conditions exist~\cite{Pioline:1998mn,GreenESeries}.

\subsection{Maximal parabolic expansions}

Let us now state the explicit expressions for the constant terms in the various maximal parabolic expansions of maximal parabolic Eisenstein series invariant under $E_9$ and $E_{10}$. The case of $E_{11}$ is treated in appendix~\ref{E11app}. In the course of this investigation, we also determine the precise numerical coefficients in the various degeneration limits. The results of this section were obtained by implementing the algorithms described in section~\ref{sec:CT} on a standard computer. We use the shorthand (\ref{maxshort}) throughout.

When writing down the expressions below one finds that for some terms it is important to consider which particular Weyl word is used to represent an element of the double coset $\mathcal{W}_{j_\circ}\backslash\mathcal{W}/\mathcal{W}_{i_*}$, appearing in the sum on the r.h.s of~(\ref{ConstMaxFin}) or~\eqref{FullConstTermAffine}. Although the sum (\ref{ConstMaxFin}) is clearly independent of the choice of representative, some Weyl words used as coset representatives can yield coefficients $M(w,\lambda)$ that appear to be infinite. In this case, the corresponding Eisenstein series goes to zero so that the product is finite. This choice of having different possible coset representatives also manifests itself in the functional relation \eqref{funcrel}. We have verified that our choice of representative gives finite Eisenstein series contributions.

\subsection*{$E_9$ Eisenstein series}

All maximal parabolic expansions of the $E_9$ Eisenstein series (\ref{e9eis}) will necessarily have two expansion parameters, namely $r$ coming from the choice of the maximal parabolic and $v$ that enters the definition (\ref{ESeriesAffine}). The additional factor of $v$ in (\ref{e9eis}) is crucial here for obtaining the right result in all cases.

\subsubsection*{Decompactification limit:}

\begin{align}
\label{e9dec1}
\int_{d+1}\mathcal{E}^{2}_{(0,0)}= r^6 v \mathcal{E}^{3}_{(0,0)}+\frac{4\zeta(6)}{3\zeta(2)}r^6v^6\,,
\end{align}
\begin{align}
\label{e9dec2}
\int_{d+1}\mathcal{E}^{2}_{(1,0)}= r^{10}v\mathcal{E}^{3}_{(1,0)}+\frac{2}{15}\zeta(2)r^{10}v^4\mathcal{E}^{3}_{(0,0)}+\frac{16\zeta(10)}{45\zeta(2)}r^{10}v^{10}\,.
\end{align}
These agree perfectly with (\ref{2ddec}) when the expansion parameters are identified as $r=\lambda$ and $v=\rho$. The final terms are consistent with the expected behaviour~\cite{GreenAutoProps,Pioline:2010kb}.

\subsubsection*{Perturbation limit:}

\begin{align}
\int_{1}\mathcal{E}^{2}_{(0,0)}= 2\zeta(3)v r^3+\frac{16}{21}\zeta(4)r^{3}E^{SO(8,8)}_{9;3}\,,
\end{align}
\begin{align}
\int_{1}\mathcal{E}^{2}_{(1,0)}= \zeta(5)v r^5+\frac{64}{297}\zeta(8)r^5E^{SO(8,8)}_{9;5}+\frac{7\zeta(6)}{3\zeta(2)}r^5v^{-1}E^{SO(8,8)}_{3;2}\,.
\end{align}
These are consistent with (\ref{2dpert}) when the expansion parameters are identified as $r=(\ell_s/\ell_2)^2$ and $v=1/y_2$. 

\subsubsection*{Semi-classical M-Theory limit:}

\begin{align}
\int_{2}\mathcal{E}^{2}_{(0,0)}=4\zeta(2)r^2v+2\zeta(3)r^2v^{2/3}E^{SL(9)}_{1;\frac32}\,,
\end{align}
\begin{align}
\int_{2}\mathcal{E}^{2}_{(1,0)}&= \zeta(5)r^{10/3}v^{4/9}E^{SL(9)}_{1;\frac52}
+\frac{4}{15}\zeta(3)\zeta(2)r^{10/3}v^{1/9}E^{SL(9)}_{3;2}\nn\\
&\quad +\frac{7\zeta(6)}{3\zeta(2)}r^{10/3}v^{10/9}E^{SL(9)}_{1;-\frac12}\,.
\end{align}
These are perfectly consistent with (\ref{2dm}) when the expansion parameters are identified as $r=(\ell_{11}/\ell_2)^3$ and $v=\mathcal{V}_9/\ell_{11}^9$.

\subsection*{$E_{10}$ Eisenstein series}

We now turn to the expansion of the $E_{10}$ Eisenstein series (\ref{e10eis}) in the three limits of section~\ref{sec:DegLimitD1}.

\subsubsection*{(Double) decompactification limit:}

Mathematically, there is no difficulty with performing the expansion of the $E_{10}$ Eisenstein series in its $E_9$ parabolic. We give the results thus obtained as well as those of an expansion in its $E_8$ parabolic, corresponding to a double decompactification.
The first $E_{10}$ Eisenstein series  (\ref{e10eis}) satisfies
\begin{align}
\label{e10dec}
\int_{10} \mathcal{E}^1_{(0,0)}
&=  v^{-1}\mathcal{E}^2_{(0,0)}+\frac{5\zeta(7)}{4\zeta(2)} v^{-7} \,,&\nn\\
\int_{10,9} \mathcal{E}^1_{(0,0)}& = a^6 \mathcal{E}^3_{(0,0)}+ \frac{4\zeta(6)}{3\zeta(2)} a^6 v^{5}+\frac{5\zeta(7)}{4\zeta(2)} v^{-7}  \,,&
\end{align}
where $a$ is the second parameter that arises in the double expansion. We see that this behaviour is consistent with (\ref{doubledec}) for $D=1$ when the expansion parameters are identified as $a=v_2$ and $v=1/r$. We also note that the single decompactification is consistent with a naive application of (\ref{decompR41}) to $D=1$ when ignoring the pre-factor. 
Performing the same analysis for the $\partial^4 R^4$ series in (\ref{e10eis}) one obtains
\begin{align}
\label{e10dec2}
\int_{10} \mathcal{E}^{1}_{(1,0)} &= v^{-1}\mathcal{E}^{2}_{(1,0)} + \frac{\zeta(5)}{4\zeta(2)} v^{-6} \mathcal{E}^{2}_{(0,0)}+\frac{7\zeta(11)}{16\zeta(2)} v^{-11} \,,&\nn\\
\int_{10,9} \mathcal{E}^{1}_{(1,0)}&= a^{10} \left(\mathcal{E}^{3}_{(1,0)} + \frac{2\zeta(2)}{15} v^{3}\mathcal{E}^{3}_{(0,0)}+ \frac{16\zeta(10)}{45\zeta(2)} v^{9}\right)&\nn\\
&\quad+ a^6 v^{-5} \left( \frac{\zeta(5)}{4\zeta(2)} \mathcal{E}^{3}_{(0,0)} + \frac{\zeta(5)\zeta(6)}{3\zeta(2)\zeta(2)} v^{5}\right)
+\frac{7\zeta(11)}{16\zeta(2)} v^{-11} \,. &
\end{align}
This is again in agreement with (\ref{doubledec}) with the same identifications as above.  However, now there is a difference that is related to the single decompactification limit: The term involving the two-dimensional $R^4$ contribution $\mathcal{E}^2_{(0,0)}$ does not appear with the right power of $v$ to be consistent with (\ref{decompR42}) without the prefactor. More precisely, the $v$ pre-factors from (\ref{decompR42}) should be $v^{-1}$, $v^{-5}$ and $v^{-11}$ rather than $v^{-1}$, $v^{-6}$ and $v^{-11}$. This cannot be compensated by the additional factor of $v$ appearing in (\ref{e9eis}) since it affects both the first two terms. It would be interesting to investigate whether this means that this particular threshold contribution in $D=2$ behaves differently from higher dimensions. We also note that the final terms are consistent with the expected behaviour~\cite{GreenAutoProps,Pioline:2010kb}.

The double decompactification in the second lines of (\ref{e10dec}) and (\ref{e10dec2}) is naturally also consistent (mathematically) with applying the $E_9$ decompactification of (\ref{e9dec1}) and (\ref{e9dec2}) to the first lines.

\subsubsection*{Perturbation limit:}
\begin{align}
\int_{1}\mathcal{E}^{1}_{(0,0)}=2\zeta(3)r^3+\frac{5\zeta(7)}{4\zeta(2)}r^{7/2}E^{SO(9,9)}_{10;\frac72}\,,
\end{align}
\begin{align}
\int_{1}\mathcal{E}^{1}_{(1,0)}= \zeta(5)r^5+\frac{7\zeta(11)}{16\zeta(2)}r^{11/2}E^{SO(9,9)}_{10;\frac{11}{2}}+\frac{7\zeta(6)}{3\zeta(2)}r^{6}E^{SO(9,9)}_{3;2}\,.
\end{align}
This is consistent with (\ref{1dpert}) for $r=y_1^2$.

\subsubsection*{Semi-classical M-Theory limit:}

\begin{align}
\int_{2}\mathcal{E}^{1}_{(0,0)}= 4\zeta(2)r^2+2\zeta(3)r^{21/10}E^{SL(10)}_{1;\frac32}\,,
\end{align}
\begin{align}
\int_{2}\mathcal{E}^{1}_{(1,0)}=\frac{7\zeta(6)}{3\zeta(2)}r^{33/10}E^{SL(9)}_{1;-\frac12}+\zeta(5)r^{7/2}E^{SL(10)}_{1;\frac52}+\frac{4}{15}\zeta(2)\zeta(3)r^{18/5}E^{SL(10)}_{3;2}\,.
\end{align}
Looking at (\ref{1dm}) we find perfect agreement for $r=(\mathcal{V}_{10}/\ell_1^{10})^{1/3}$.

In summary, we have found that our proposals (\ref{e9eis}) and (\ref{e10eis}) for the $E_9$ and $E_{10}$ Eisenstein series are consistent with the physical conditions that we deduced above and the evaluation of the maximal parabolic expansions also provided the precise numerical coefficients in the various relations. Before subjecting the functions to further checks we make some more general remarks on the number of constant terms in the minimal parabolic expansion the structure of the Eisenstein series.

\subsection{Minimal parabolic expansion}
\label{sec:minimalexpansion}

Now consider the minimal parabolic expansion of maximal parabolic Eisenstein series. The explicit expressions for the minimal parabolic expansions of $E_{1;\frac32}^G$ and $E_{1;\frac52}^G$ with $G=E_9$, $E_{10}$ and $E_{11}$ can be found in appendix~\ref{sec:expansion}. These expressions are directly obtained by evaluating~\eqref{FullConstTermAffine} (without $v$ for $E_{10}$ and $E_{11}$). We stress that these series develop logarithmic and (logarithm)$^2$ terms from taking limits in the $\xi$-functions entering $M(w,\lambda)$.

In a general expansion of $E_{1;s}^G$, it is instructive to count the number of Weyl words in the sum on the r.h.s of~\eqref{FullConstTermAffine}, for which the corresponding factors $M(w,\lambda)$ are non-vanishing (but possibly infinite). We do this for a range of values of the parameter $s$ and for the $E_{n\geq6}$ groups, i.e., in dimensions $0\leq D\leq5 $. The results are shown in Table \ref{svalues} which shows the number of contributing Weyl words as a function of $s$ for the various $E_n$. This is evaluated in the normalisation of the Eisenstein series $E_{1;s}^{E_n}$ that we have been using throughout the paper. Let us explain some of the structure found in this table. Our explanations make use of the $E_n$ root system and are specific to this series. 

\begin{table}
\begin{center} 
\scalebox{.88}{
\begin{tabular}{ | c | c  c  c  c  c  c  c  c  c  c  c  c  c  c | }
 \hline&&&&&&&&&&&&&& \\[-3mm]                      
  $s$ & $0$ & $\frac12$ & $1$ & $\frac32$ & $2$ & $\frac52$ & $3$ & $\frac72$ & $4$ & $\frac92$ & $5$ & $\frac{11}{2}$ & $6$ & $\frac{13}{2}$ \\[1mm] \hline\hline
 $E_6$ & $1$ & $2$ & $27$ & $7$ & $12$ & $27$ & $\cdots$ & $$ & $$ & $$ & $$ & $$ & $$ &  \\ \hline
 $E_7$ & $1$ & $2$ & $126$ & $8$ & $14$ & $35$ & $56$ & $126$ & $91$ & $126$ & $\cdots$ & $$ & $$ &  \\ \hline
 $E_8$ & $1$ & $2$ & $2160$ & $9$ & $16$ & $44$ & $72$ & $408$ & $534$ & $1060$ & $1460$ & $1795$ & $2160$ & $\cdots$ \\ \hline
 $E_9$ & $1$ & $2$ & $\infty$ & $10$ & $18$ & $54$ & $90$ & $\infty$ & $\cdots$ & $$ & $$ & $$ & $$ &  \\ \hline
 $E_{10}$ & $1$ & $2$ & $\infty$ & $11$ & $20$ & $65$ & $110$ & $\infty$ & $\cdots$ & $$ & $$ & $$ & $$ &  \\ \hline
 $E_{11}$ & $1$ & $2$ & $\infty$ & $12$ & $22$ & $77$ & $132$ & $\infty$ & $\cdots$ & $$ & $$ & $$ & $$ &  \\ \hline
\end{tabular}}
\end{center}
\caption{\em The table shows the number of Weyl words with non-vanishing coefficients $M(w,\lambda)$ in an expansion of $E^{E_{d+1}}_{1;s}$ in dimensions $1\leq D\leq5 $ and for a range of values for the parameter $s$. The ellipsis signifies that the row is continued with the last number explicitly written out (for $D\leq2$ this is conjectural). 
\label{svalues}}
\end{table}

\begin{itemize}
\item For $s=0$ there is only the identity Weyl word yielding a non-vanishing $M(w,\lambda)$ factor.
\vskip0.5mm
For $s=0$, $\langle\lambda|\alpha\rangle=-\text{ht}(\alpha)<0$ (where $\alpha>0$) and therefore there cannot be any $c(+1)$ (infinite) factors appearing in $M(w,\lambda)$. On the other hand, the simple root $\alpha_{i_*}$, for which $\langle\lambda|\alpha_{i_*}\rangle=-1$, will be included in the product for all $w\in\mathcal{S}_{i_*}^{(\infty)}$, apart from the identity element $\id$. Hence all elements, except for $\id$,  of $\mathcal{S}_{i_*}^{(\infty)}$ will yield factors of $M(w,\lambda)$ which vanish, due to presence of at least one $c(-1)$ (zero) factor and no $c(+1)$ factor to cancel it. The value of $M(\id,-\rho)=1$ and in fact the whole Eisenstein is a constant equal to $1$.

\item For $s=1/2$ there are two Weyl words producing (potentially) non-vanishing $M(w,\lambda)$ factors.
\vskip0.5mm
It is easy to see that there does not exist a root $\alpha$ such that $\langle\lambda|\alpha\rangle=1$ and therefore there will also not be $c(+1)$ factors in any of the $M(w,\lambda)$.
The first few Weyl words in $S_{i_*}^{(\infty)}$ are $\{\id, w_{i_*},w_k w_{i_*}, ...\}$, where $w_k$ is the fundamental Weyl reflection corresponding to some node $k$ connected to the node of $\alpha_{i_*}$ in the Dynkin diagram. However, the word $w_kw_{i_*}$ turns the root $\alpha_{i_*}+\alpha_k$ into a negative root and the inner product of this root with $\lambda$ is equal to $-1$, hence producing a factor of $c(-1)$ in $M(w,\lambda)$. Therefore the only contributing Weyl words for $s=1/2$ are the identity $\id$ and $w_{i_*}$, giving rise to two contributions to the constant term. In fact, they are of opposite sign but same magnitude and also the moduli dependence is the same such that the two terms cancel. The Eisenstein series has completely vanishing constant term and in fact is zero. This is due to our normalisation (\ref{minimalparabfinite}) and well-known for $SL(2,\mathbb{Z})$. Multiplying by the (string theory) normalising factor $2\zeta(2s)$ will render this Eisenstein series finite with two constant terms, one polynomial and one logarithmic.\footnote{We note that the normalisation of Eisenstein series is a subtle issue in general. For example, for $D_m$, $E_6$, $E_7$ and $E_8$ this was addressed in~\cite{GRS} where specific normalisations were derived that were shown to have as residues at $s=3/2$ the minimal representation  associated  with these Eisenstein series. Changing the normalisation also has an effect on the functional relation (\ref{funcrel}), see the end of this section~\ref{sec:minimalexpansion}. Since the numbers in table~\ref{svalues}  are related to the normalisation (\ref{minimalparabfinite}) that we are using, they do not necessarily reflect the simpler functional relation of \cite{GRS}; taking the limit to these specific $s$-values can result in several Weyl words coalescing to the same constant term.}

\item For $s=1$ the number of Weyl words with non-vanishing $M(w,\lambda)$ is given by the number of elements in the Weyl orbit $\mathcal{O}_{i_*}$.
\vskip0.5mm
We have the product $\langle\alpha_{i_*}|\lambda\rangle=1$ and $\alpha_{i_*}$ is mapped to a negative root by all Weyl elements needed for the Weyl orbit $\mathcal{O}_{i_*}$. The simple root $\alpha_{i_*}$ therefore gives an infinite $c(+1)$ contribution in the product of $M(w,\lambda)$. The only roots that have an inner product with $\lambda$ that is equal to $-1$ are the simple roots other than $\alpha_{i_*}$. However these simple roots are never mapped to a negative root due to the defining property of $\mathcal{S}_{i_*}^{(\infty)}$ and hence there are no zero $c(-1)$ factors in $M(w,\lambda)$. Therefore all Weyl elements of the orbit $\mathcal{O}_{i_*}$ contribute. The Eisenstein series is in the (degenerate) principal representation.

\item Some systematics can also be found in the columns of other values of $s$. The number appearing in a particular row can be related to the rank $n$ of the corresponding $E_n$ group. We make the following simple observations:
\begin{itemize}
\item[-] column $s=3/2$ for $E_n$ is given by $n+1$. This is equivalent to adding the numbers $1$ and $n$ which are the entries for $E_{n-1}$ for $s=0$ and $s=3/2$; this is a reflection of the decompactification rule (\ref{decompR41}). [They correspond to the trivial and minimal representation contributing to the 1/2 BPS curvature correction term.]
\item[-] column $s=2$ is given by $2n$.
\item[-] the numbers in the column with $s=5/2$ for $E_n$ are obtained by adding the values $1$, $n$ and the $s=5/2$ value for $E_{n-1}$. This is a reflection of the decompactification rule (\ref{decompR42}). [They correspond to the trivial, minimal and next-to-minimal representation contributing to the 1/4 BPS curvature correction term.]
\item[-] column $s=3$ is given by $n(n+1)$.\footnote{We thank J. Dillies for pointing this out to us.} There are also sum rules for the even values of $s$ similar to those for $s=3/2$ and $s=5/2$.
\end{itemize}
\end{itemize}

We also observe that for example in the row of $E_7$ the numbers are not always increasing when the value of $s$ is increased. For example, when going from $s=3/2$ to $s=4$, the numbers decrease from $126$ to $91$. 

For the finite-dimensional groups it is clear that when increasing the value of $s$, one will eventually always reach a threshold value. For larger values of $s$ the number of Weyl words yielding non-vanishing $M(w,\lambda)$ factors will always be equal to the dimension of the Weyl orbit $\mathcal{O}_{i_*}$. The reason for this is that for large enough values of $s$ no positive root $\alpha$ exists which satisfies $\langle\alpha|\lambda\rangle=-1$. Hence all possible terms will be present in the sum over elements of $\mathcal{S}_{i_*}$. For the infinite-dimensional groups the situation regarding this issue is less clear, since for these groups there are roots of arbitrary height available. In a sporadic check for some values of $s\geq7/2$, the calculation on a computer of the constant term did not terminate within a reasonably short period of time (in contrast with the computations for $s<7/2$). This can be taken as an tentative indication that in these cases the number of Weyl words contributing is actually infinite.\footnote{Physically, this may be related to curvature correction terms unprotected by supersymmetry.} This is the reason why we put $\infty$ for the corresponding entries in Table \ref{svalues}.

Looking at Table \ref{svalues} it is tempting to interpret it as a strong sign for the special properties associated with the small values of $s$ in the set 
\begin{align}
s\in\left\{0,1/2,1,3/2,2,5/2,3\right\}\,.
\end{align}
More precisely, by requiring the constant term to only encode a \textit{finite} number of perturbative effects as required by supersymmetry, the range of possible values that $s$ can take, gets reduced from a previously infinite set to a finite number of possible values. 
It would be desirable to make these statements more precise and to prove them rigorously.

We finally also note that the series $E^{E_n}_{1;s}$ for $n=6,7,8$ in the normalisation (\ref{minimalparabfinite}) have the following simple poles and simple zeroes in $s$ in the `critical strips' defined below: 
\begin{itemize}
\item $E_6$: zeroes at $s=\frac12,2$; poles at $s=\frac92,6$.
\item $E_7$: zeroes at $s=\frac12,2,3,\frac{17}4$; poles at $s=\frac92,6,7,\frac{17}2$.
\item $E_8$: zeroes at $s=\frac12,2,3,\frac72,\frac{17}4,\frac{23}4$; poles at $s=6,\frac{13}2,7,\frac{15}2,\frac{17}2,9,10,\frac{23}2$.
\end{itemize}
If the we change the normalisation of the $E_6$ and $E_7$ series according to \cite{GRS,Pioline:2010kb} and suitably for $E_8$, the zeroes disappear and the simple poles lie at
\begin{itemize}
\item $\xi(2s)\xi(2s-3) E_{1;s}^{E_6}$: poles at $s=0,\frac32,\frac92,6$.
\item $\xi(2s)\xi(2s-3)\xi(2s-5)\xi(4s-16) E_{1;s}^{E_7}$: poles at $s=0,\frac32,\frac52,4,\frac92,6,7,\frac{17}2$.
\item $\xi(2s)\xi(2s-3)\xi(2s-5)\xi(2s-6)\xi(2s-9)\xi(4s-16)\xi(4s-22) E_{1;s}^{E_8}$: poles at $s=0,\frac32,\frac52,3,4,\frac92,5,\frac{11}2,6,\frac{13}2,7,\frac{15}2,\frac{17}2,9,10,\frac{23}2$.
\end{itemize}
For $E_7$, this is consistent with the functional relation~\cite{GRS,Pioline:2010kb} that relates $s\leftrightarrow \frac{17}2-s$. For $E_6$ one has $s\leftrightarrow 6-s$ but one also has to change the weight from node $1$ to node $5$ by going to the contragredient representation~\cite{GRS,Pioline:2010kb}. We have verified these functional relations explicitly on the constant terms in a number of examples from the list above. The list of poles is complete in the intervals $[0,6]$ for $E_6$ and $[0,17/2]$ for $E_7$. 
The normalising factor for the $E_{1;s}^{E_8}$ series is  such that the resulting completed Eisenstein series is symmetric under $s\leftrightarrow \frac{23}2-s$.\footnote{For completeness, we also give the pole structure of another completed Eisenstein series that was discussed in~\cite{GRS,Pioline:2010kb}. This is the series associated with node $8$ of the $E_8$ diagram. The normalising factor for $E_{8;s}^{E_8}$ is $\xi(2s)\xi(2s-5)\xi(2s-9)\xi(4s-28)$ and the remaining simple poles lie at $s=0,\frac52,\frac92,7,\frac{15}2,10,12,\frac{29}2$. The completed maximal parabolic Eisenstein series associated with this node is invariant under $s\leftrightarrow\frac{29}2-s$.} The list of poles for $E_8$ is complete in the interval $[0,23/2]$. Note that the normalising factor has a double pole at $s=3$ so that the single zero in the unnormalised series is turned into a (simple) pole in the completed Eisenstein series.

In all cases, the residues of the completed Eisenstein series at $s=\frac32$ correspond to the minimal and at $s=\frac52$ to the next-to-minimal representation~\cite{GRS,Pioline:2010kb,GreenSmallRep}.  The value $s=0$ gives the trivial representation. It would be interesting to know if the residues for larger values of $s$ are related to other special (small) automorphic representations.

\subsection{Laplace eigenvalues}
\label{sec:Laplace}

We now  perform another consistency check on the Eisenstein series (\ref{e9eis})--(\ref{e11eis}) that derives from their Laplace eigenvalues which were already mentioned in the introduction. Let us first state the form of these equations, which can also be found in \cite{GreenAutoProps}. The (almost) homogeneous Laplace eigenvalue equations, satisfied by the first two coefficients are for $D\geq 3$
\begin{align}\label{evalue00}
\left(\Delta^{D}-\frac{3(11-D)(D-8)}{D-2}\right)\mathcal{E}^{D}_{(0,0)}&=6\pi\delta_{D,8}\,,&\\
\label{evalue10}
\left(\Delta^{D}-\frac{5(12-D)(D-7)}{D-2}\right)\mathcal{E}^{D}_{(1,0)}&=40\zeta(2)\delta_{D,7}\,.&
\end{align}
The inhomogeneous Laplace equation for the $\mathcal{E}^{D}_{(0,1)}$ coefficient takes the form
\begin{align}\label{evalue01}
\left(\Delta^{D}-\frac{6(14-D)(D-6)}{D-2}\right)\mathcal{E}^{D}_{(0,1)}=-\left(\mathcal{E}^{D}_{(0,0)}\right)^2+120\zeta(3)\delta_{D,6}\,.
\end{align}
Here the $\delta_{i,j}$ are discrete Kronecker deltas and $\Delta^{D}$ is the Laplace operator defined on $\mathcal{M}_{d+1}$ where $d=10-D$. These were derived in~\cite{GreenAutoProps} using the decompactification limit of the Laplace operator from $D$ to $D+1$ dimensions. We see that for $D=2$  all three equations appear to break down. This is an artefact of the method of derivation that needs to be refined for $D=2$ as we already saw in section~\ref{sec:DegLimitD2} when studying the decompactification limit. Performing the analysis more carefully\footnote{In the metric (\ref{met2d}) the circle direction $\rho$ does not have a quadratic kinetic term, it is associated with a light-like direction. This changes the structure of the relation between the Laplace operators $\Delta^2$ and $\Delta^3$ compared to in higher dimensions. More precisely, we have that in the decompactification limit
\begin{align}
\Delta^2 \to \Delta^3 - 30 \rho\partial_\rho\nn
\end{align}
in terms of the metric component $\rho=r_d/\ell_3$ of (\ref{met2d}), leading to the eigenvalues in (\ref{2dev}). Recall that we chose to keep $\lambda$ fixed in the decompactification limit such that no derivatives with respect to $\lambda$ appear. The coefficient $30$ is related to the volume of moduli space including all dual potentials in $D=2$ and is equal to $\frac12\cdot 2\cdot h^\vee$, where $h^\vee$ is the dual Coxeter number of $E_9$ that also appears in the shift of the standard affine Weyl vector~\cite{Kac90}. Incidentally, it agrees also with the linear term in the analysis of~\cite{GreenAutoProps}.}
 using the metric (\ref{met2d}) one arrives at the following equations in $D=2$ for $R^4$, $\partial^4 R^4$ and $\partial^6 R^4$
\begin{align}
\label{2dev}
\left(\Delta^{2}+150\right)\mathcal{E}^{2}_{(0,0)}&=0\,,\nn\\
\left(\Delta^{2}+210\right)\mathcal{E}^{2}_{(1,0)}&=0\,,\nn\\
\left(\Delta^{2}+228\right)\mathcal{E}^{2}_{(0,1)}&=-\left(\mathcal{E}^{2}_{(0,0)}\right)^2\,.
\end{align}
(We recall that our definition of the $D=2$ Laplacian was explained in section~\ref{sec:EKM}.) Below $D=2$ the Laplace eigenvalues obey again (\ref{evalue00})--(\ref{evalue01}).

These eigenvalues can also be obtained from the quadratic Casimir evaluated for the weights $-(\lambda+\rho)$ defining the Eisenstein series. Before going into a discussion of the specific case of $D=2$ however, let us explain in some detail the relation between the Laplace operator $\Delta^D$ and the quadratic Casimir $\Omega$ of the corresponding duality algebra in $D$ dimensions. The following results were obtained in very useful discussions with H. Nicolai.

The general definition of the quadratic Casimir is
\begin{align}
\Omega_\Lambda=\langle\Lambda+2\rho|\Lambda\rangle\,,
\end{align}
on a highest weight representation with highest weight $\Lambda$ and $\rho$ is the Weyl vector of the algebra, see \cite{Kac90}. To see how this expression connects with the eigenvalues given above, let us evaluate $\Omega_\Lambda$ for various choices of $\Lambda$.\\
If we choose $\Lambda=-2k\Lambda_2$, then the evaluated Casimir expression is
\begin{align}
\Omega_{-2k\Lambda_2}=\frac{2k(11-D)}{D-2}(2k+3D-26)\,,
\end{align}
where $\Lambda_2$ is the fundamental weight associated with the exceptional node in $D$ dimensions. Specialising to $k=1$ leaves us with
\begin{align}
\Omega_{-2\Lambda_2}=\frac{6(11-D)(D-8)}{D-2}\,,
\end{align}
which, when divided by $2$, is the eigenvalue appearing in the equation \eqref{evalue00}.\footnote{In general one always has to compare the eigenvalues $\frac12\Omega=\Delta^D$.} Using the weight $-2\Lambda_2$ is Weyl equivalent to using $-3\Lambda_1$, which upon evaluation of the quadratic Casimir yields the same eigenvalue (see e.g.~\cite{ObersStringThresh}). This weight is also precisely the weight used to define the coefficient Eisenstein series of $\mathcal{R}^4$.

Let us comment on the motivation for picking $\Lambda=-2k\Lambda_2$ here. The motivation for this particular choice comes from the BKL analysis carried out in \cite{CurvCorrDamour,CurvCorrHanany}. In the BKL analysis, which was first proposed in \cite{Belinski} for gravity in four dimensions and then extended to higher-dimensional and supersymmetric theories in \cite{DH01,NicDamHenn}  one makes the BKL-like ansatz
\begin{align}
ds_{10}^2=e^{2\alpha\phi}ds^2_D+\sum_{a=1}^{11-D}e^{-2\beta^a}\theta^a\otimes\theta^a
\end{align}
for a metric in $10$ dimensions, with the triangular frame $\theta^a=\mathcal{N}^a_jdx^j$. The $\beta^a$ appearing in the metric are also the variables, which parameterise the Cartan subalgebra of the $E_{d+1}$ algebra.

Then in \cite{CurvCorrDamour,CurvCorrHanany}, various curvature corrections were analysed and the corresponding dominant BKL walls were calculated. For instance it was found that the dominant wall corresponding to a correction of the form $\mathcal{R}^{1+3k}$ descending from $D=11$ is given by $k\Lambda_2$. The term in the BKL Lagrangian then is of the form
\begin{align}
\mathcal{L}_{\text{BKL}}\sim e^{-2k\Lambda_2(\beta^a)}\,.
\end{align}
Following the idea that there is a close connection between the BKL analysis and the relevant curvature corrections in string theory, we arrive at the particular choice of $\Lambda$ made above. The `BKL wall' in the preceding equation is like the dilaton pre-factors one obtains from a toroidal compactification and hence is a term in the constant term of the duality invariant curvature correction term. If one assumes Weyl invariance of the constant terms like one has for (\ref{minparexp}) the quadratic Casimir can be evaluated on any piece of the constant term. Therefore the calculation of the quadratic Casimir has to give the same result as the Laplacian. A Weyl equivalent representative of $-2\Lambda_2$ is $-3\Lambda_1$, the weight more commonly used for the $R^4$ curvature correction term and the one that we have used throughout the preceding sections.

The eigenvalues of the coefficients of the next two higher orders in curvature corrections can also be reproduced by evaluating the quadratic Casimir for specific weights. For the $\partial^4R^4$ correction a weight that reproduces the eigenvalue in \eqref{evalue10} is $-5\Lambda_1$, corresponding to the $s=5/2$ case.  We remark that in the  $\partial^6R^4$ case one can for example reproduce the `eigenvalue' of the inhomogeneous Laplace equation (\ref{evalue01}) by using the BKL wall weight $\frac{d+4}2\Lambda_{d+1}$ for $E_{d+1}$ (hence $\lambda=(d+4)\Lambda_{d+1}-\rho$). However, this weight does not have a Weyl equivalent representative that uses a dominant combination of $\Lambda_1$ and $\Lambda_2$ and it is thus hard to imagine how it could arise from a type II string calculation.\footnote{Possible combinations of $\Lambda_1$ and $\Lambda_2$ for BKL walls in $D\neq 2$ are $-\frac32\Lambda_1+3\Lambda_2$ and $\frac{15}4\Lambda_1-\frac12\Lambda_2$.}

Let us now consider the specific case of $D=2$, i.e the case where $\Omega$ is the quadratic Casimir of the full affine $E_9$ algebra. When we evaluate the quadratic Casimir for the weights derived from the values for $\hat{\lambda}$ given in (\ref{e9eis}) one recovers the values in (\ref{2dev}). Again, it is important to use the weight that is shifted by $\hat{\delta}$ otherwise one obtains the wrong value.

The quadratic Casimir values for $E_{10}$ and $E_{11}$ again agree with those of (\ref{evalue00})--(\ref{evalue10}). There are no subtleties in these cases as the rank of $E_{d+1}$ equals the number of circle directions.

\section{Conclusion}

In the present paper we have considered the perturbative sector of type II  superstring four-graviton scattering amplitudes in $D$ dimensions that are expected to be invariant under the discrete $E_{d+1}$ duality groups of table~\ref{dualitygroups} with $d=10-D$. We have extended existing results for the finite-dimensional duality groups to the infinite-dimensional duality groups $E_9$, $E_{10}$ and $E_{11}$. This was done by defining Eisenstein series, which are the coefficients of the few lowest orders in the scattering amplitude. It was found that for special choices of the parameter $s$ which appears in the definition of these Eisenstein series, the part in the expansion of the Eisenstein series which corresponds to the constant term, contains a finite number of terms. We considered the expansion of these Eisenstein series in terms of its minimal parabolic subgroup and in terms of three different maximal parabolic subgroups corresponding to the physical degeneration limits discussed in section~\ref{sec:computations}. In all cases, we found exact agreement with the assumption that the constant terms of the respective Eisenstein series encode a finite number of perturbative string theory corrections, namely the lowest few loop-orders of the scattering process. In the course of demonstrating this agreement we also presented a careful analysis of the limits in two-dimensions where many formula from higher dimensions superficially appear to break down. As noted below (\ref{e10dec}), the $E_{10}$ series (\ref{e10eis}) for $D=1$ is consistent with the double decompactification limit (\ref{doubledec}) to $D=3$ but predicts a single decompactification limit to $D=2$ that differs in one of the threshold terms from the general pattern.

The values  $s=0$, $s=3/2$ and $s=5/2$ that appear are quite special and are related to small automorphic representations being associated with the Eisenstein series defined for these values~\cite{Kazhdan:2001nx,Pioline:2010kb,GreenSmallRep,GRS} in the case of finite-dimensional duality groups. From the dramatic collapse of the constant terms from a generic infinite number to a small finite number at these values it appears natural to propose that also here there are small automorphic representations underlying these particular Eisenstein series also in the Kac--Moody case. This is something that might be possible to check by a further detailed analysis of the abelian and non-abelian Fourier coefficients, something that is beyond the scope of this paper. In a similar vein, it would be most interesting to have a description of these series in terms of (constrained) lattice sums that exhibit the BPS states that contribute.

We emphasise that irrespective of their actual occurrence in scattering amplitudes in very low space-time dimensions, the Eisenstein series for Kac--Moody groups considered in this paper provide an economical tool for summarizing the automorphic functions that are relevant for $R^4$ and $\partial^4 R^4$ curvature correction terms in higher dimensions. These can be obtained by expanding the constant terms in smaller parabolic subgroups than maximal ones. One example considered in the main text was the double decompactification of the $E_{10}$ series; a different example involving $E_{11}$ giving rise to the $E_7$ series is considered in the appendix.

Finally, we note that automorphic forms for $E_{10}(\mathbb{Z})$ and for the Weyl group of $E_{10}$ have appeared in different conjectures concerning M-theory~\cite{Ganor:1999ui,Brown:2004jb} and quantum gravity~\cite{Kleinschmidt:2009cv,Kleinschmidt:2010zz} and we hope that our investigations can prove useful for these ideas. It would also be interesting to see how much can be learned about curvature correction terms with more space-time derivatives, starting from $\partial^6 R^6$ where the automorphic function is not expected to be a pure Eisenstein series since the Laplace equation is inhomogeneous~\cite{GreenAutoProps,Pioline:2010kb,GreenESeries}.


\subsection*{Acknowledgements}

The authors are grateful to T.~Damour, S.~Fredenhagen, M.~Green, H.~Nicolai, S.~Miller, D.~Persson, B.~Pioline and  P.~Vanhove for useful discussions and correspondence. We would also like to thank D.~Persson for comments on a draft of this paper and B.~Pioline for detailed comments on the first version.
The work of PF is supported by the Erasmus Mundus Joint Doctorate Program by Grant Number 2010-1816 from the EACEA of the European Commission and the Universit\'e de Nice-Sophia Antipolis. AK would also like to thank the Isaac Newton Institute, Cambridge, for hospitality while part of this work was carried out and the programme organisers there for providing a stimulating work environment.

\appendix

\section{$E_{11}$ maximal parabolic expansions}
\label{E11app}

In this appendix, we give for completeness the maximal parabolic expansions of the $E_{11}$ Eisenstein series (\ref{e11eis}) using the shorthand (\ref{maxshort}).

\subsubsection*{Decompactification limit}

The decompactification limits corresponding to the Levi factor $GL(1)\times E_{10}$ for the $s=3/2$ and $s=5/2$ series are
\begin{align}
\int_{11} \mathcal{E}_{(0,0)}^0 &= r^{-6} \mathcal{E}_{(0,0)}^1 + \frac{12\zeta(6)}{5\pi} r^8\,,&\nn\\
\int_{11} \mathcal{E}_{(1,0)}^0 &= r^{-10} \mathcal{E}_{(1,0)}^1 +\frac{2\zeta(6)}{3\pi\zeta(2)} \mathcal{E}_{(0,0)}^1 + \frac{16\zeta(12)}{9\pi\zeta(2)} r^{12}\,.
\end{align}

The powers of $r$ and the structure of the resulting Eisenstein series are in agreement with (\ref{declim}) applied naively to $D=0$  when one replaces the `$0$-dimensional Planck length' $\ell_0$ by the radius of the first direction and $\ell_1$ according to the standard Kaluza--Klein rules. The final terms are consistent with the expected behaviour~\cite{GreenAutoProps,Pioline:2010kb}.

\subsubsection*{Perturbative limit}

\begin{align}
\int_{1} \mathcal{E}_{(0,0)}^0 &= 2\zeta(3)r^3+\frac{12\zeta(6)}{5\pi}r^4 E^{SO(10,10)}_{11;4} \,,&\nn\\
\int_{1} \mathcal{E}_{(1,0)}^0 &= \zeta(5) r^5
 + \frac{16\zeta(12)}{9\pi\zeta(2)} r^{6}  E^{SO(10,10)}_{11;6} 
 +\frac{4\zeta(4)}{3}r^7 E^{SO(10,10)}_{3;2} \,.
\end{align}
The  powers of $r$ and the structure of the $SO(10,10)$ Eisenstein series are in agreement with the naive application of (\ref{pertlim}).

\subsubsection*{Semi-classical M-theory limit}

\begin{align}
\int_{2} \mathcal{E}_{(0,0)}^0 &= 4\zeta(2)r^2+r^{24/11} E^{SL(11)}_{1;3/2} \,,&\nn\\
\int_{2} \mathcal{E}_{(1,0)}^0 &=  \frac{4\zeta(4)}{3}r^{36/11} E^{SL(11)}_{1;1/2}
 + \zeta(5)  r^{40/11}  E^{SL(11)}_{1;5/2} 
 +\frac{4\zeta(2)}{15}\zeta(3) r^{42/11} E^{SO(10,10)}_{3;2} \,.
\end{align}
The  powers of $r$ and the structure of the $SL(11)$ Eisenstein series are in agreement with the naive application of (\ref{r4m1}) and (\ref{r4m2}).

\subsubsection*{Four-dimensional limit}

As a final application, we consider the Levi decomposition of $E_{11}$ with Levi factor $SL(4)\times GL(1)\times E_7$ as appropriate for an interpretation in $D=4$. This corresponds to removing node $8$ from the Dynkin diagram. Expanding the constant terms of the Eisenstein series (\ref{e11eis}) under the associated maximal parabolic one obtains\footnote{While this work was being completed, the preprint~\cite{Gubay:2012sy} appeared that also studies parameters related to `middle' nodes of the $E_n$ diagram  (like our $r$ here) and deduces the first terms in our two expansions (\ref{e7e11}).}
\begin{align}
\label{e7e11}
\int_{8} \mathcal{E}_{(0,0)}^0 &= r^3 \mathcal{E}_{(0,0)}^4 + \frac{3\zeta(5)}{\pi} r^2 E^{SL(4)}_{9;-2}\,,\nn\\
\int_{8} \mathcal{E}_{(1,0)}^0 &=  r^5 \mathcal{E}_{(1,0)}^4 
+ \frac{\zeta(3)}{\pi} r^{9/2} E^{SL(4)}_{9;-1} \mathcal{E}_{(0,0)}^4
+\frac{\pi\zeta(5)}{15} r^{7/2} E^{SL(4)}_{10;-3/2}\nn\\
&\quad + \frac{7\zeta(9)}{12\pi} r^3 E^{SL(4)}_{9;-2}
 \,.
\end{align}
Here, $r=\left(\text{vol}(T^4)\ell_0^8/\ell_4^4\right)^{1/3}$ parameterises the $GL(1)$ factor in the Levi part as usual and the (maximal) Eisenstein series on the r.h.s. belong to $SL(4)\times E_7$ and we have factorized them. Note again, that our (non-standard) labelling for $E_n$ subgroups is obtained from diagram~\ref{fig:Edplus1Diag} by removing nodes. Here, this means that $SL(4)$ inherits the three nodes labelled $11$, $10$ and $9$ while $E_7$ has nodes $1$ up to $7$. The leading terms are the pure $E_7$ Eisenstein series as they appear in $D=4$ and we have used the relation (\ref{ESeries}).

The constant terms of the $SL(4)$ Eisenstein series can now be analysed in their minimal parabolic, leaving only dependence on four dilatonic scalars (including $r$) and $E_7$ Eisenstein series. Then one sees more clearly the expected feature that the $E_{11}$ series knows about the relevant series in $D=4$ but also about threshold contributions. As always with derivative corrections the term with the highest number of derivatives (here $\partial^4 R^4$) in $D$ dimensions induces the terms with up to that number of derivatives in higher space-time dimensions. In this sense, going to higher rank $E_n$ groups combines the information of derivative corrections of different orders in single objects. (This was stressed to us by P. Vanhove.)

\section{Minimal parabolic expansions}\label{sec:expansion}

In this appendix, we give the minimal parabolic expansions of the $E_9$, $E_{10}$ and $E_{11}$ maximal parabolic Eisenstein series with $s=3/2$ and $s=5/2$. Note that in each case the Eisenstein series which we expand do not include the additional normalisation factors of $2\zeta(3)$ and $\zeta(5)$ shown in \eqref{e9eis}--\eqref{e11eis}.\\
In the expressions below, $\gamma_{\text{E}}$ is the Euler-Mascheroni constant and $A$ denotes the Glaisher-Kinkelin constant. We note that the `number' of terms here does not need to strictly agree with table~\ref{svalues} since taking the limits to $s=3/2$ and $s=5/2$ in the factors $M(w,\lambda)$ can produce several terms out of a single Weyl word $w$. The first terms in all expressions is that of the identity Weyl word and corresponds to the string perturbation tree level term.

The variables $r_i$ in the expressions below are defined by parameterising the Cartan subalgebra via a basis of simple roots. More precisely, we let the function $H$ of (\ref{Hfunc}) be $H(a)=\sum_{i=1}^r r_i \alpha_i$, where $r$ is the rank of the algebra (excluding the derivation in the $E_9$ case) and $\alpha_i$ are the simple roots. With this choice, the (minimal parabolic) constant term of the maximal parabolic Eisenstein series with weight $\lambda=2s\Lambda_1-\rho$ starts out with $r_1^{2s}$. $r_1$ is the string coupling, the other $r_i$ are different combinations of the physical parameters.

\subsection*{$E_9$ Eisenstein series}

The constant terms of the maximal parabolic Eisenstein series $E^{E_9}_{1;3/2}$ in the minimal parabolic read
\begin{align}
&r_1^3+\frac{r_6^3}{r_7^2}+\frac{ \pi ^3 r_7^4}{45 r_8^3 \zeta(3)}+ \frac{2 \pi \gamma_{\text{E}} r_4 }{\zeta(3)}-\frac{2\pi r_4\log(4 \pi  r_5)}{\zeta(3)}+\frac{4\pi r_4 \log(r_4)}{\zeta(3)}-\frac{2\pi r_4\log(r_3)}{\zeta(3)}\nn\\
&+\frac{2\pi r_4\log(r_2)}{\zeta(3)}+ \frac{\pi^2r_5^2}{3r_6}+\frac{\pi^2r_3^2}{3r_1}+\frac{\pi^2r_2^2}{3}+\frac{4 \pi ^4 r_9^6}{945 v^5 \zeta(3)}+\frac{3 r_8^5 \zeta(5)}{2 \pi  r_9^4 \zeta(3)}
\end{align}

The constant terms of the maximal parabolic Eisenstein series $E^{E_9}_{1;5/2}$ in the minimal parabolic read
\begin{align}
&r_1^5+\frac{\pi  r_4^5}{15 r_5^4}+\frac{\pi  r_8^5}{15 v^3}-\frac{4 \zeta(3) \log(r_7) r_6r_1^3}{\zeta(5)}+\frac{2 \zeta(3)^2 r_4^3r_1^3}{\pi  r_3^2 r_2^2 \zeta(5)}+\frac{4 \pi ^7 r_7^8}{70875 r_8^7 \zeta(5)}+\frac{2 \pi ^3 r_7^2 r_5^2}{9 r_8 r_6 \zeta(5)}\nn\\
&+\frac{8 \pi ^6 r_5^6}{42525 r_6^5 \zeta(5)}+\frac{2 \pi ^4 r_3^4}{135 r_1^3 \zeta(5)}+\frac{2 \pi ^3 r_7^2 r_3^2}{9 r_8 r_1 \zeta(5)}-\frac{4 \pi ^2 \log(r_7) r_6 r_3^2}{3 r_1 \zeta(5)}+\frac{2 \pi ^3 r_5^2 r_3^2}{9 r_4 r_1 \zeta(5)}\nn\\
&+\frac{2 \pi ^3 r_7^2 r_2^2}{9 r_8 \zeta(5)}+\frac{2 \pi ^3 r_5^2 r_2^2}{9 r_4 \zeta(5)}+\frac{2 \pi ^5 r_3^4 r_2^4}{2025 r_4^3 \zeta(5)}+\frac{2 \pi ^4 r_2^4}{135 r_1 \zeta(5)}+\frac{32 \pi ^8 r_9^{10}}{1403325 v^9 \zeta(5)}\nn\\
&+\frac{2 \pi ^5 r_7^4 r_9^4}{2025 r_8^3 v^3 \zeta(5)}+\frac{2 \pi ^4 r_5^2 r_9^4}{135 r_6 v^3 \zeta(5)}+\frac{2 \pi ^4 r_3^2 r_9^4}{135 r_1 v^3 \zeta(5)}+\frac{2 \pi ^4 r_2^2 r_9^4}{135 v^3 \zeta(5)}+\frac{2 \pi  r_6^3 \zeta(3)}{3 r_8 \zeta(5)}\nn\\
&+\frac{2 \pi  r_4^3 \zeta(3)}{3 r_3^2 \zeta(5)}+\frac{2 \pi  r_7^2 r_1^3 \zeta(3)}{3 r_8 \zeta(5)}+\frac{2 \pi  r_5^2 r_1^3 \zeta(3)}{3 r_4 \zeta(5)}+\frac{2 \pi ^2 r_3^4 \zeta(3)}{45 r_2^2 \zeta(5)}+\frac{2 \pi  r_4^3 \zeta(3)}{3 r_1 r_2^2 \zeta(5)}+\frac{2 \pi ^2 r_1^3 r_2^4 \zeta(3)}{45 r_3^2 \zeta(5)}\nn\\
&+\frac{2 \pi ^2 r_7^4 \zeta(3)}{45 r_9^2 \zeta(5)}+\frac{2 \pi  r_8^3 r_5^2 \zeta(3)}{3 r_6 r_9^2 \zeta(5)}+\frac{2 \pi  r_8^3 r_3^2 \zeta(3)}{3 r_1 r_9^2 \zeta(5)}+\frac{2 \pi  r_8^3 r_2^2 \zeta(3)}{3 r_9^2 \zeta(5)}+\frac{2 \pi ^2 r_6^3 r_9^4 \zeta(3)}{45 r_7^2 v^3 \zeta(5)}\nn\\
&+\frac{2 \pi ^2 r_1^3 r_9^4 \zeta(3)}{45 v^3 \zeta(5)}+\frac{2 r_8^3 r_6^3 \zeta(3)^2}{\pi  r_7^2 r_9^2 \zeta(5)}+\frac{2 r_8^3 r_1^3 \zeta(3)^2}{\pi  r_9^2 \zeta(5)}+\frac{r_6^7 \zeta(7)}{6 r_7^6 \zeta(5)}+\frac{7 r_8^9 \zeta(9)}{12 \pi  r_9^8 \zeta(5)}\nn\\
&+\frac{4 \pi ^2r_6 r_3^2}{3 r_1 \zeta(5)}\Big(\gamma_{\text{E}}+2 \log(r_6)-\log(4 \pi  r_5\Big)+\frac{4r_6 r_1^3 \zeta(3)}{\zeta(5)} \Big(\gamma_{\text{E}}+2 \log(r_6)-\log(4 \pi  r_5)\Big)\nn\\
&+\frac{4 \pi ^2r_6 r_2^2}{3 \zeta(5)} \Big(\gamma_{\text{E}}-\log(4 \pi  r_7)+2 \log(r_6)-\log(r_5)\Big)\nn\\
&+\frac{4 \pi ^3r_4 r_9^4}{45 v^3 \zeta(5)} \Big(\gamma_{\text{E}}-\log(4 \pi  r_5)+2 \log(r_4)-\log(r_3)-\log(r_2)\Big)\nn\\
&+\frac{4 \pi ^2r_7^2 r_4}{3 r_8 \zeta(5)}\Big( \gamma_{\text{E}}-\log(4 \pi  r_5)+2 \log(r_4)-\log(r_3)-\log(r_2)\Big)\nn\\
&+\frac{4 \pi ^2r_5^2}{3 \zeta(5)}\Big(2 \gamma_{\text{E}}-24 \log(A)-\log(r_7)+2 \log(r_5)-\log(r_3)-\log(r_2)\Big)\nn\\
&+\frac{4r_8^3 r_4 \zeta(3)}{r_9^2 \zeta(5)} \Big(\gamma_{\text{E}}-\log(4 \pi  r_5)+2 \log(r_4)-\log(r_3)-\log(r_2)\Big)\nn\\
&+\frac{8 \pi r_6 r_4}{\zeta(5)} \Big(\gamma_{\text{E}}^2-4 \gamma_{\text{E}} \log(2)+4 \log(2)^2+\log(\pi )^2+2 (-\gamma_{\text{E}}+\log(4)) \log(\pi  r_5)\nn\\
&\quad+\log(r_7) (-\gamma_{\text{E}}+\log(4 \pi  r_5))+2 (\gamma_{\text{E}}-\log(4 \pi  r_7)) \log(r_4)\nn\\
&\quad+2 \log(r_6) (\gamma_{\text{E}}-\log(4 \pi  r_5)+2 \log(r_4)-\log(r_3)-\log(r_2))\nn\\
&\quad-(\gamma_{\text{E}}-\log(4 \pi  r_7))(\log(r_3)+\log(r_2))+\log(r_5) (2 \log(\pi )+\log(r_5)\nn\\
&\quad-2 \log(r_4)+\log(r_3)+\log(r_2))\Big)
\end{align}

\subsection*{$E_{10}$ Eisenstein series}

The constant terms of the maximal parabolic Eisenstein series $E^{E_{10}}_{1;3/2}$ in the minimal parabolic read
\begin{align}
&r_1^3+\frac{r_6^3}{r_7^2}+\frac{4 \pi^4 r_9^6}{945 r_{10}^5\zeta(3)}+\frac{\pi^3 r_7^4}{45 r_8^3\zeta(3)}+\frac{\pi^2 r_5^2}{3 r_6\zeta(3)}+\frac{2 \pi \gamma_{\text{E}} r_4}{\zeta(3)}+\frac{\pi^2  r_3^2}{3 r_1 \zeta(3)}-\frac{2 \pi r_4\log(4 \pi  r_5)}{ \zeta(3)}\nn\\
&+\frac{4 \pi r_4\log(r_4)}{\zeta(3)}-\frac{2 \pi r_4\log(r_3)}{\zeta(3)}-\frac{2 \pi r_4\log(r_2)}{\zeta(3)}+\frac{\pi^2 r_2^2}{3 \zeta(3)}+\frac{3  r_8^5 \zeta(5)}{2 \pi r_9^4 \zeta(3)}+\frac{15 r_{10}^7 \zeta(7)}{4 \pi^2 \zeta(3)}
\end{align}

The constant terms of the maximal parabolic Eisenstein series $E^{E_{10}}_{1;5/2}$ in the minimal parabolic read
\begin{align}
&r_1^5+\frac{4}{315} \pi ^2 r_9^6+\frac{\pi  r_8^5}{15 r_{10}^3}+\frac{\pi  r_{10}^5 r_7^4}{15 r_8^3}+\frac{r_{10}^5 r_5^2}{r_6}+\frac{\pi  r_4^5}{15 r_5^4}+\frac{r_{10}^5 r_3^2}{r_1}+r_{10}^5 r_2^2+\frac{3 r_{10}^5 r_6^3 \zeta(3)}{\pi ^2 r_7^2}\nn\\
&+\frac{3 r_{10}^5 r_1^3 \zeta(3)}{\pi ^2}+\frac{2 \pi ^5 r_9^4 r_7^4}{2025 r_{10}^3 r_8^3 \zeta(5)}+\frac{4 \pi ^7 r_7^8}{70875 r_8^7 \zeta(5)}-\frac{4 \zeta(3) \log(r_7) r_6 r_1^3}{\zeta(5)}+\frac{2\zeta(3)^2 r_4^3 r_1^3}{\pi  r_3^2 r_2^2 \zeta(5)}\nn\\
&+\frac{32 \pi ^8 r_9^{10}}{1403325 r_{10}^9 \zeta(5)}+\frac{2 \pi ^4 r_9^4 r_5^2}{135 r_{10}^3 r_6 \zeta(5)}+\frac{2 \pi ^3 r_7^2 r_5^2}{9 r_8 r_6 \zeta(5)}+\frac{8 \pi ^6 r_5^6}{42525 r_6^5 \zeta(5)}+\frac{2 \pi  r_6^3 \zeta(3)}{3 r_8 \zeta(5)}\nn\\
&+\frac{2 \pi ^4 r_3^4}{135 r_1^3 \zeta(5)}+\frac{2 \pi ^4 r_9^4 r_3^2}{135 r_{10}^3 r_1 \zeta(5)}+\frac{2 \pi ^3 r_7^2 r_3^2}{9 r_8 r_1 \zeta(5)}-\frac{4 \pi ^2 \log(r_7) r_6 r_3^2}{3 r_1 \zeta(5)}+\frac{2 \pi ^3 r_5^2 r_3^2}{9 r_4 r_1 \zeta(5)}\nn\\
&+\frac{2 \pi ^4 r_9^4 r_2^2}{135 r_{10}^3 \zeta(5)}+\frac{2 \pi ^2 r_9^4 r_1^3 \zeta(3)}{45 r_{10}^3 \zeta(5)}+\frac{2 \pi  r_7^2 r_1^3 \zeta(3)}{3 r_8 \zeta(5)}+\frac{2 \pi ^3 r_7^2 r_2^2}{9 r_8 \zeta(5)}+\frac{r_6^7 \zeta(7)}{6 r_7^6 \zeta(5)}+\frac{7 r_8^9 \zeta(9)}{12 \pi  r_9^8 \zeta(5)}\nn\\
&+\frac{21 r_{10}^{11} \zeta(11)}{8 \pi ^2 \zeta(5)}+\frac{2 \pi ^3 r_5^2 r_2^2}{9 r_4 \zeta(5)}+\frac{2 \pi ^5 r_3^4 r_2^4}{2025 r_4^3 \zeta(5)}+\frac{2 \pi ^4 r_2^4}{135 r_1 \zeta(5)}+\frac{2 \pi ^2 r_7^4 \zeta(3)}{45 r_9^2 \zeta(5)}+\frac{2 \pi ^2 r_9^4 r_6^3 \zeta(3)}{45 r_{10}^3 r_7^2 \zeta(5)}\nn\\
&+\frac{2 \pi  r_8^3 r_5^2 \zeta(3)}{3 r_9^2 r_6 \zeta(5)}+\frac{2 \pi  r_4^3 \zeta(3)}{3 r_3^2 \zeta(5)}+\frac{2 \pi  r_8^3 r_3^2 \zeta(3)}{3 r_9^2 r_1 \zeta(5)}+\frac{2 \pi  r_5^2 r_1^3 \zeta(3)}{3 r_4 \zeta(5)}+\frac{2 \pi ^2 r_3^4 \zeta(3)}{45 r_2^2 \zeta(5)}+\frac{2 \pi  r_4^3 \zeta(3)}{3 r_1 r_2^2 \zeta(5)}\nn\\
&+\frac{2 \pi  r_8^3 r_2^2 \zeta(3)}{3 r_9^2 \zeta(5)}+\frac{2 \pi ^2 r_1^3 r_2^4 \zeta(3)}{45 r_3^2 \zeta(5)}+\frac{2 r_8^3 r_6^3 \zeta(3)^2}{\pi  r_9^2 r_7^2 \zeta(5)}+\frac{2 r_8^3 r_1^3 \zeta(3)^2}{\pi  r_9^2 \zeta(5)}+\frac{9 r_{10}^5 r_8^5 \zeta(5)}{2 \pi ^3 r_9^4}\nn\\
&+\frac{4 r_6 r_1^3 \zeta(3)}{\zeta(5)} \Big(\gamma_{\text{E}}+2 \log(r_6)-\log(4 \pi  r_5)\Big)+\frac{4 \pi ^2  r_6 r_3^2}{3 r_1 \zeta(5)}\Big(\gamma_{\text{E}}+2 \log(r_6)-\log(4 \pi  r_5)\Big)\nn\\
&+\frac{4r_8^3 r_4 \zeta(3)}{r_9^2 \zeta(5)}\Big(\gamma_{\text{E}}-\log(4 \pi  r_5)+2 \log(r_4)-\log(r_3)-\log(r_2)\Big)\nn\\
&+\frac{4 \pi ^2 r_6 r_2^2}{3 \zeta(5)} \Big(\gamma_{\text{E}}-\log(4 \pi  r_7)+2 \log(r_6)-\log(r_5)\Big)\nn\\
&+\frac{4 \pi ^3r_9^4 r_4}{45 r_{10}^3 \zeta(5)}\Big(\gamma_{\text{E}}-\log(4 \pi  r_5)+2 \log(r_4)-\log(r_3)-\log(r_2)\Big)\nn\\
&+\frac{4 \pi ^2r_7^2 r_4}{3 r_8 \zeta(5)} \Big(\gamma_{\text{E}}-\log(4 \pi  r_5)+2 \log(r_4)-\log(r_3)-\log(r_2)\Big)\nn\\
&+\frac{6r_{10}^5 r_4}{\pi } \Big(\gamma_{\text{E}}-\log(4 \pi  r_5)+2 \log(r_4)-\log(r_3)-\log(r_2)\Big) \nn\\
&+\frac{4 \pi ^2r_5^2}{3 \zeta(5)} \Big(2 \gamma_{\text{E}}-24 \log(A)-\log(r_7)+2 \log(r_5)-\log(r_3)-\log(r_2)\Big)\nn\\
&+\frac{8 \pi r_6 r_4}{\zeta(5)} \Big(\gamma_{\text{E}}^2-4 \gamma_{\text{E}} \log(2)+4 \log(2)^2+\log(\pi )^2+2 (-\gamma_{\text{E}}+\log(4)) \log(\pi  r_5)\nn\\
&\quad+\log(r_7) (-\gamma_{\text{E}}+\log(4 \pi  r_5))+2 (\gamma_{\text{E}}-\log(4 \pi  r_7)) \log(r_4)\nn\\
&\quad+2 \log(r_6) (\gamma_{\text{E}}-\log(4 \pi  r_5)+2 \log(r_4)-\log(r_3)-\log(r_2))\nn\\
&\quad-(\gamma_{\text{E}}-\log(4 \pi  r_7))(\log(r_3)+\log(r_2))\nn\\
&\quad+\log(r_5) (2 \log(\pi )+\log(r_5)-2 \log(r_4)+\log(r_3)+\log(r_2))\Big)
\end{align}

\subsection*{$E_{11}$ Eisenstein series}

The constant terms of the maximal parabolic Eisenstein series $E^{E_{11}}_{1;3/2}$ in the minimal parabolic read
\begin{align}
&r_1^3+\frac{r_6^3}{r_7^2}+\frac{2 \pi ^5 r_{11}^8}{1575\zeta(3)}+\frac{4 \pi ^4 r_9^6}{945\zeta(3) r_{10}^5}+\frac{\pi ^3 r_7^4}{45\zeta(3) r_8^3}+\frac{\pi ^2 r_5^2}{3\zeta(3) r_6}+\frac{2 \gamma_{\text{E}} \pi  r_4}{\zeta(3)}+\frac{ \pi^2  r_3^2}{3\zeta(3) r_1}\nn\\
&+\frac{\pi^2  r_2^2}{3\zeta(3)}+\frac{3 r_8^5 \zeta(5)}{2 \pi  r_9^4 \zeta(3)}+\frac{15 r_{10}^7 \zeta(7)}{4 \pi ^2 r_{11}^6 \zeta(3)}
-\frac{6\pi r_4\log(4 \pi  r_5)}{3\zeta(3)}+\frac{12 \pi r_4\log(r_4)}{3\zeta(3)}\nn\\
&-\frac{6 \pi r_4\log(r_3)}{3\zeta(3)}-\frac{6\pi r_4\log(r_2)}{3\zeta(3)}
\end{align}

The constant terms of the maximal parabolic Eisenstein series $E^{E_{11}}_{1;5/2}$ in the minimal parabolic read
\begin{align}
&r_1^5+\frac{4 \pi ^2 r_9^6}{315 r_{11}^4}+\frac{\pi  r_8^5}{15 r_{10}^3}+\frac{4 \pi ^2 r_{11}^6 r_8^5}{315 r_9^4}+\frac{\pi  r_{10}^5 r_7^4}{15 r_{11}^4 r_8^3}+\frac{r_{10}^5 r_5^2}{r_{11}^4 r_6}+\frac{\pi  r_4^5}{15 r_5^4}+\frac{r_{10}^5 r_3^2}{r_{11}^4 r_1}+\frac{r_{10}^5 r_2^2}{r_{11}^4}\nn\\
&+\frac{3 r_{10}^5 r_6^3 \zeta(3)}{\pi ^2 r_{11}^4 r_7^2}+\frac{3 r_{10}^5 r_1^3 \zeta(3)}{\pi ^2 r_{11}^4}
-\frac{4 \zeta(3) \log(r_7) r_6 r_1^3}{\zeta(5)}+\frac{2 \zeta(3)^2 r_4^3 r_1^3}{\pi \zeta(5) r_3^2 r_2^2}\nn\\
&-\frac{4 \pi ^2 \log(r_7) r_6 r_3^2}{3 r_1 \zeta(5)}+\frac{22112 \pi ^9 r_{11}^{12}}{1915538625 \zeta(5)}+\frac{32 \pi ^7 r_{11}^6 r_9^6}{893025 r_{10}^5 \zeta(5)}+\frac{32 \pi ^8 r_9^{10}}{1403325 r_{10}^9 \zeta(5)}\nn\\
&+\frac{8 \pi ^6 r_{11}^6 r_7^4}{42525 r_8^3 \zeta(5)}+\frac{2 \pi ^5 r_9^4 r_7^4}{2025 r_{10}^3 r_8^3 \zeta(5)}+\frac{4 \pi ^7 r_7^8}{70875 r_8^7 \zeta(5)}+\frac{2 \pi  r_7^2 r_1^3 \zeta(3)}{3 r_8 \zeta(5)}+\frac{8 \pi ^5 r_{11}^6 r_5^2}{2835 r_6 \zeta(5)}\nn\\
&+\frac{2 \pi ^4 r_9^4 r_5^2}{135 r_{10}^3 r_6 \zeta(5)}+\frac{2 \pi ^3 r_7^2 r_5^2}{9 r_8 r_6 \zeta(5)}+\frac{8 \pi ^6 r_5^6}{42525 r_6^5 \zeta(5)}+\frac{2 \pi ^4 r_3^4}{135 r_1^3 \zeta(5)}+\frac{8 \pi ^5 r_{11}^6 r_3^2}{2835 r_1 \zeta(5)}\nn\\
&+\frac{2 \pi ^4 r_9^4 r_3^2}{135 r_{10}^3 r_1 \zeta(5)}+\frac{2 \pi ^3 r_7^2 r_3^2}{9 r_8 r_1 \zeta(5)}
+\frac{2 \pi ^3 r_5^2 r_3^2}{9 r_4 r_1 \zeta(5)}+\frac{8 \pi ^5 r_{11}^6 r_2^2}{2835 \zeta(5)}+\frac{2 \pi ^4 r_9^4 r_2^2}{135 r_{10}^3 \zeta(5)}+\frac{2 \pi ^3 r_7^2 r_2^2}{9 r_8 \zeta(5)}\nn\\
&+\frac{2 \pi ^3 r_5^2 r_2^2}{9 r_4 \zeta(5)}+\frac{2 \pi ^5 r_3^4 r_2^4}{2025 r_4^3 \zeta(5)}+\frac{2 \pi ^4 r_2^4}{135 r_1 \zeta(5)}+\frac{2 \pi ^2 r_7^4 \zeta(3)}{45 r_9^2 \zeta(5)}+\frac{2 \pi  r_6^3 \zeta(3)}{3 r_8 \zeta(5)}+\frac{8 \pi ^3 r_{11}^6 r_6^3 \zeta(3)}{945 r_7^2 \zeta(5)}\nn\\
&+\frac{2 \pi ^2 r_9^4 r_6^3 \zeta(3)}{45 r_{10}^3 r_7^2 \zeta(5)}+\frac{2 \pi  r_8^3 r_5^2 \zeta(3)}{3 r_9^2 r_6 \zeta(5)}+\frac{2 \pi  r_4^3 \zeta(3)}{3 r_3^2 \zeta(5)}+\frac{2 \pi  r_8^3 r_3^2 \zeta(3)}{3 r_9^2 r_1 \zeta(5)}+\frac{8 \pi ^3 r_{11}^6 r_1^3 \zeta(3)}{945 \zeta(5)}\nn\\
&+\frac{2 \pi ^2 r_9^4 r_1^3 \zeta(3)}{45 r_{10}^3 \zeta(5)}+\frac{2 \pi  r_5^2 r_1^3 \zeta(3)}{3 r_4 \zeta(5)}+\frac{2 \pi ^2 r_3^4 \zeta(3)}{45 r_2^2 \zeta(5)}+\frac{2 \pi  r_4^3 \zeta(3)}{3 r_1 r_2^2 \zeta(5)}+\frac{2 \pi  r_8^3 r_2^2 \zeta(3)}{3 r_9^2 \zeta(5)}\nn\\
&+\frac{2 \pi ^2 r_1^3 r_2^4 \zeta(3)}{45 r_3^2 \zeta(5)}+\frac{2 r_8^3 r_6^3 \zeta(3)^2}{\pi  r_9^2 r_7^2 \zeta(5)}+\frac{2 r_8^3 r_1^3 \zeta(3)^2}{\pi  r_9^2 \zeta(5)}+\frac{9 r_{10}^5 r_8^5 \zeta(5)}{2 \pi ^3 r_{11}^4 r_9^4}+\frac{2 \pi  r_{10}^7 \zeta(7)}{63 \zeta(5)}\nn\\
&+\frac{r_6^7 \zeta(7)}{6 r_7^6 \zeta(5)}+\frac{4 r_8^3 r_4 \zeta(3)}{r_9^2 \zeta(5)} \Big(\gamma_{\text{E}}-\log(4 \pi  r_5)+2 \log(r_4)-\log(r_3)-\log(r_2)\Big)\nn\\
&+\frac{7 r_8^9 \zeta(9)}{12 \pi  r_9^8 \zeta(5)}+\frac{4 \pi ^2r_6 r_2^2}{3 \zeta(5)} \Big(\gamma_{\text{E}}-\log(4 \pi  r_7)+2 \log(r_6)-\log(r_5)\Big)\nn\\
&+\left(\frac{4 r_6 r_1^3 \zeta(3)}{\zeta(5)} +\frac{4 \pi ^2 r_6 r_3^2}{3 r_1 \zeta(5)}\right)\Big(\gamma_{\text{E}}+2 \log(r_6)-\log(4 \pi  r_5)\Big)\nn\\
&+\frac{4 \pi ^2 r_5^2}{3 \zeta(5)} \Big(2 \gamma_{\text{E}}-24 \log(A)-\log(r_7)+2 \log(r_5)-\log(r_3)-\log(r_2)\Big)\nn\\
&+\frac{21 r_{10}^{11} \zeta(11)}{8 \pi ^2 r_{11}^{10} \zeta(5)}+\frac{6r_{10}^5 r_4}{\pi  r_{11}^4} \Big(\gamma_{\text{E}}-\log(4 \pi  r_5)+2 \log(r_4)-\log(r_3)-\log(r_2)\Big) \nn\\
&+\frac{16 \pi ^4r_{11}^6 r_4}{945 \zeta(5)} \Big(\gamma_{\text{E}}-\log(4 \pi  r_5)+2 \log(r_4)-\log(r_3)-\log(r_2)\Big) \nn\\
&+\frac{4 \pi ^3 r_9^4 r_4}{45 r_{10}^3 \zeta(5)} \Big(\gamma_{\text{E}}-\log(4 \pi  r_5)+2 \log(r_4)-\log(r_3)-\log(r_2)\Big)\nn\\
&+\frac{4 \pi ^2r_7^2 r_4}{3 r_8 \zeta(5)} \Big(\gamma_{\text{E}}-\log(4 \pi  r_5)+2 \log(r_4)-\log(r_3)-\log(r_2)\Big) \nn\\
&+\frac{8 \pi r_6 r_4}{\zeta(5)} \Big(\log(r_7) (-\gamma_{\text{E}}+\log(4 \pi  r_5))+4 \log(2)^2+2 (-\gamma_{\text{E}}+\log(4)) \log(\pi  r_5)\nn\\
&\quad \gamma_{\text{E}}^2-4 \gamma_{\text{E}} \log(2)+2 \log(r_6) (\gamma_{\text{E}}-\log(4 \pi  r_5)+2 \log(r_4)-\log(r_3)-\log(r_2))\nn\\
&\quad+2 (\gamma_{\text{E}}-\log(4 \pi  r_7)) \log(r_4)-(\gamma_{\text{E}}-\log(4 \pi  r_7))(\log(r_3)+\log(r_2))\nn\\
&\quad+\log(\pi )^2+\log(r_5) (2 \log(\pi )+\log(r_5)-2 \log(r_4)+\log(r_3)+\log(r_2))\Big)
\end{align}

{\small
\bibliography{KacMoodyGravScatterv2}
\bibliographystyle{utphys}
}

\end{document}